\documentclass[preprint,onecolumn,showpacs,superscriptaddress,nofootinbib]{revtex4-2}
 \usepackage[cp1251]{inputenc}
 \usepackage[figurename=Fig.]{caption}
\usepackage[tablename=Table ]{caption}
\usepackage{graphicx}\usepackage{epsfig}
\usepackage[psamsfonts]{amssymb}
\usepackage{amsfonts}\usepackage{amscd}
\usepackage{amsmath,amssymb}
\usepackage{cases}
\usepackage{relsize}

\newcommand{\ba}{\begin{eqnarray}}
\newcommand{\ea}{\end{eqnarray}}

\def \Imm { \mbox{\rm Im} }
\def \Ree { \mbox{\rm Re} }
\def \XD {{\cal{X}}}
\def \Res { \mbox{\rm Res} }
\def \aE {a_E}

\usepackage{graphicx}\usepackage{epsfig}
\usepackage{amsfonts}\usepackage{amscd}
\usepackage{amsmath,amssymb}
\everymath{\displaystyle}

\graphicspath{{./Figures-mixed/}}
\usepackage{color}

\begin{document}

 \title{The sixth order QED radiative  corrections to    lepton
anomalies due to  the fourth order vacuum polarization insertions within the Mellin-Barnes representation}

\author{ L.P. Kaptari }
\email{kaptari@theor.jinr.ru} \affiliation{Bogoliubov Lab. Theor.
Phys., JINR, Dubna, 141980, Russia}
\author{V.I. Lashkevich}
\email{lashkevich@gstu.gomel.by} \affiliation{Gomel State Technical
University, Gomel, 246746, Belarus}
\author{O.P. Solovtsova}
\email{solovtsova@gstu.gomel.by; olsol@theor.jinr.ru}
\affiliation{Bogoliubov Lab. Theor. Phys., JINR, Dubna, 141980,
Russia}
 \affiliation{Gomel State Technical
University, Gomel, 246746, Belarus}

 \begin{abstract}
The explicit form of the sixth order radiative  corrections   to the lepton  $L$ ($L=e, ~\mu $ and $\tau$)
  anomalous magnetic moment from QED Feynman diagrams   with insertion of the fourth-order
   polarization operators  consisting of either two closed lepton loops or one lepton loop
   crossed by a photon line is discussed in detail. The approach is based on the consistent
   application of dispersion relations for vacuum polarization operators and the Mellin-Barnes
   transform for massive photon propagators. Explicit analytical expressions for  corrections to
   the lepton anomaly are obtained  for the first time  in the whole interval $0 <r< \infty $ of
   the ratio $r$ of lepton masses $m_\ell/m_L$. Asymptotic expansions in the limit of both small
    $r\ll 1$ and large $r\gg 1$  computed from the exact expressions   are found to be in  perfect
     agreement with the ones earlier reported in the literature.  We argue that in the region where
      the physical lepton mass ratios are located, the asymptotic expansions hold with an accuracy
      higher than the experimentally measured anomalies. The two loop diagrams with all three leptons
       different from each other   are computed numerically and compared with the corresponding
       corrections from the pure two-bubble  and  one-bubble mixed diagrams. It is shown that there
       are regions of ratios $r$ where all three types of the fourth order polarization operator  contribute
       equally to the anomaly.
 \end{abstract}


\date{\today}

 \maketitle

\section{Introduction}
The electron and muon magnetic moments are now one of the most precisely measured quantities in particle
 physics and allows one to thoroughly test the relativistic local Quantum Field Theory (QFT) with tremendous accuracy. The physics
of the  gyromagnetic factor $g_{L}$ defined as the ratio between the magnetic moment $\mu $  and the spin $s$
 of a lepton $L$ with charge $e$ and mass $m$ ($\vec\mu =g_L\displaystyle\frac{e }{2m}\vec s$)  has long
 challenged the particle physics community, and experiments and theory   look rather intricate.
 Its significance is tightly connected with its early role as a valuable test of QFT. Dirac’s
  relativistic theory of quantum mechanics remarkably predicted that the $g_L$-factor of a
  free point-like fermion should be exactly~2, see Ref.~\cite{dirac}. However, later on J.~Schwinger~\cite{Schwinger1948}
  computed the leading-order quantum correction to the lepton magnetic moment arising from the self-interaction with
  a virtual photon and found a small correction $\sim \alpha/2\pi$, where $\alpha$ is the electromagnetic
  fine structure constant, which explained the unexpected $\sim 0.12\% $ excess observed in precision
   measurements of the electron magnetic moment, a discrepancy referred to as the anomalous magnetic
    moment of the electron. This  success served as an additional  argument in the foundation of
    Quantum Electrodynamics (QED) as the true,  authentic theory of electromagnetic interactions and QFT
    as a general framework for the theory of elementary particles.

It is now conventional to define the anomalous magnetic moment of a fermion $L$ as $a_L = (g_L -2)/2$  which quantifies the deviation of $g_L$  from the Dirac value of 2. Since then, the anomalous magnetic moment has been measured with incredible accuracy becoming the best measured quantity in the Standard Model (SM) and allows one to test relativistic local QFT  with unprecedented accuracy. It imposes severe limits on deviations from the standard theory of elementary particles; therefore, investigations of the anomalous magnetic moments have challenged the particle physics community for a long time, and experiments and theory look hitherto highly topical. The electron and muon magnetic moment provide the most precise tests of QED in particular and of relativistic local QFT as a common framework for elementary particle theory in general.

Presently, the electron and muon   $g$-factors are known with an accuracy   $\sim$0.13 ppt~\cite{Fan:2022eto} and $\sim$ 124 ppb~\cite{Muong-2:2025xyk}, respectively. This corresponds to a relative accuracy of $\sim 10^{-10}$  for electrons  and $\sim 10^{-6}$
 for muons. A comprehensive analysis  of   different mechanisms to $a_\mu$    within the Standard Model (SM) can be found in, e.g.,  Ref.~\cite{Aliberti:2025beg}; for an overview of the current status from   both, the experimental and theoretical points of view,  see Ref.~\cite{Gabrielse:2025jep}. The measurements of the electron and muon $g$-factor along with the corresponding theoretical calculations are also aimed at probing the limits of the SM or the possible existence of as yet undiscovered
  particles Beyond the Standard Model (BSM). In this context,
  the muon anomaly $a_\mu$ serves as  an extremely powerful tool in investigations of the BSM  physics, q.v. Refs.~\cite{Athron:2025ets,Pustyntsev:2025nwm}.

In the SM, the lepton anomaly $a_{L}$ is calculated from the perturbative expansion in the fine-structure constant $\alpha$. The most important contributions to $\alpha_L$  arise from pure QED diagrams, but starting from the order $\alpha^2$, hadronic contributions, in particular hadron vacuum polarization and hadron light-by-light scattering, also become notable.
Nowadays, high precision calculations of QED corrections (up to the fifth order in $\alpha$) are performed by means of special computational algorithms, e.g., high-precision calculation  by difference equations~\cite{Laporta2000,LaportaPLB} augmented by the PSLQ-algorithm~\cite{Ferguson}. These methods assure several hundred or even thousand significant digits in the final results~\cite{Laporta2000,LaportaPLB}. In spite of the ability to perform so high precision  calculations,  these approaches turn  to be quite cumbersome in practice application. Even in the case of parallel computation realisation, one requires a huge amount of computer time. Moreover, due to cumbersomeness of the method, a detailed analysis of the results with a separate investigation of the role of a particular type of diagrams, as well as independent confirmation  of the numerical results, appear to be rather awkward.
In this context, it is quite tempting  to identify
a subgroup of specific diagrams, which enable  close  analytical calculations, in the full set of diagrams  of a given  order. This subgroup consists exclusively of diagrams with insertions of the vacuum photon polarization operator with the inclusion of self-energy corrections of the corresponding order. The simplest type of such  operators is the one  consisting solely  of  closed lepton loops, usually referred to as the ``bubble''-like diagrams.
Based on the Mellin-Barnes representation for the massive propagators, cf. Refs.~\cite{Davych,Friot:2005cu,Friot:2011ic}  (a generalization of the Mellin-Barnes technique to the multifold  Mellin-Barnes integrals can be found in  Refs.~\cite{FriotBanik1,FriotBanik2}),
a detailed analytical consideration of this type of diagrams was presented  Ref.~\cite{Aguilar:2008qj}. It was shown that
the application of the Mellin–Barnes approach makes it possible to derive general expressions for the corrections to the muon $a_{\mu}$ in the form of one (two)-dimensional integrals which determine the Mellin momenta of the corresponding polarization operator.
In Ref.~\cite{Solovtsova23}, this approach was generalized to the case of  any arbitrary lepton $L=e,
~\mu$ and $\tau$, $a_{L}$, and for the whole interval of the mass ratio $r=\,m_\ell/m_L\,$ of the loop lepton $\ell$ to the external $L$ one
in the whole region $0 < r < \infty $.
More recent applications of the Mellin-Barnes approach to $a_L$ can be found in Refs.~\cite{Charles:2017snx,Ananthanarayan:2020acj,Solovtsova-2024}.

The next in complexity operators are diagrams with closed lepton loops crossed by photon lines, hereafter referred to as the ``mixed''\ type.
This is a  natural extension of our earlier calculations~\cite{Solovtsova23,Solovtsova-2024} of corrections from  the
  bubble-like diagrams,
by including in the consideration the  mixed diagrams.
Obviously, the lowest order of these diagrams is $\alpha^2$   that contributes to the sixth- and higher-order radiative corrections to $a_L$. In this paper, we mainly focus on analytical calculations of the sixth order corrections stipulated by accounting for  only one mixed loop in  the vacuum polarization operators.
As before, our consideration is based on a combined use of the dispersion relations for the polarization operators and the Mellin-Barnes integral transform for the Feynman parametric integrals.
Notice that the very first  analytical expressions  for mixed diagrams were reported  in Ref.~\cite{Laporta:1993ju}   in the asymptotic limits  $ r\ll 1 $ and  $ r\gg 1 $. Herebelow,  our efforts are aimed at exact calculations of the mixed diagrams for any value of $r$, $0 < r < \infty $.

Our paper is organized as follows. In Section~II,  the most general expressions for the  vertex radiative corrections from diagrams with vacuum polarization insertions  are obtained within the Mellin-Barnes approach. Explicit formulae for the corresponding Mellin momenta are presented for an arbitrary number of lepton loops, for both, the bubble-like and the mixed types of diagrams. The sixth order of corrections to $a_L$ are analysed  analytically in Section~III. The Subsection~\ref{direct} is devoted to numerical calculation of integrals by the direct formula. The corrections to $a_L$, computed  analytically by the Cauchy theorem for the right and left semi-planes, are presented in Subsections~\ref{right} and \ref{left}, respectively. Numerical analysis of the obtained analytical formulae is presented in Section~\ref{num}. In this section we also discuss the asymptotic behaviour  at large $r\gg 1$ and small $r\ll 1$
and the range of applicability of the asymptotic formulae.
The contributions from mixed and bubble diagrams
are discussed in subsection~\ref{bubb}.
Conclusions are collected in Section~\ref{concl}.
Some important explanatory  formulae
are delegated to Appendices~\ref{appA}~and~\ref{appB}.

\section{Vacuum polarization  and  lepton anomaly}

In  this section we present the most general expressions for the  vertex radiative corrections from diagrams with vacuum polarization insertions. A particular case when the vacuum polarization operator consists only of closed lepton loops was previously considered in detail in Refs.~\cite{Aguilar:2008qj,Solovtsova23,Solovtsova-2024}. Here below we consider a more complicated polarization operator which includes also loops with internal photon lines.
The renormalized photon propagator    can be written as

\vskip 0.6cm
 \begin{figure}[h]
 \phantom{}\hspace*{-0.8cm}%
\includegraphics[width=0.98\textwidth,clip=true]{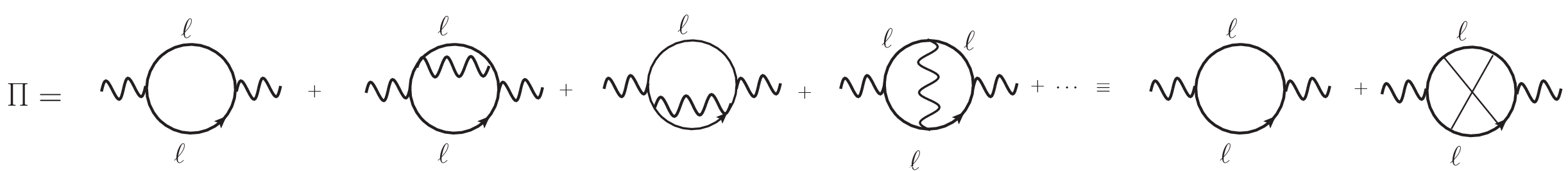} \caption{The irreducible proper graphs contributing to the vacuum
polarization operator up to the fourth order. The crossed loop is the shorthand notation for the  sum of diagrams with one internal
photon line.}
\label{Irpolar}
 \end{figure}
\ba && D_{\alpha\beta}(k^2) =  -i\left\{
\left( g_{\alpha\beta}-\frac{k_\alpha k_\beta}{k^2}  \right)\frac{1}{k^2} \dfrac{1}{1+     \mathlarger{ \mathlarger{ \Pi (k^2)}}}+\xi\frac{k_\alpha k\beta}{k^4}\right\} =\nonumber\\
&& -i\left\{ \left(g_{\alpha\beta}-\frac{k_\alpha k_\beta}{k^2}\right) \frac{1}{k^2}\left[ 1-\left( \Pi (k^2)-\Pi^2(k^2)+\Pi^3(k^2)+\cdots\right)\right]+\xi\frac{k_\alpha k\beta}{k^2} \right\}
\equiv\nonumber \\ &&
-i\left\{\left( g_{\alpha\beta}-\frac{k_\alpha k_\beta}{k^2}  \right)\frac{1}{k^2}\left[ 1- \widetilde\Pi (k^2)\right]+\xi\frac{k_\alpha k\beta}{k^4}\right\} , \label{prop} \ea
where $\xi$ is the gauge fixing parameter and $ \Pi(k^2)$ is the sum of all   irreducible self-energy graphs which,  up to the fourth order, are depicted in
Fig.~\ref{Irpolar}. Since the sum of   irreducible diagrams remains transverse, the longitudinal part of the propagator
does not affect our calculations and, consequently,  we can choose the full propagator as pure transverse. This
 corresponds to the choice $\xi =0$, i.e., to  the  Landau gauge. Moreover, it can be shown that the remaining part proportional
 to $k_\alpha k_\beta$ also does not contribute to the anomaly $a_L$ and, consequently, in our calculations we can restrict ourselves   to only terms the
 $\sim g_{\alpha\beta}$, see   Refs.~\cite{Solovtsova23,lautrupNuclPhys,Aguilar:2008qj}  and Appendix~\ref{appA}.

\begin{figure}[h]
\phantom{}\vspace{0.8cm}%
\includegraphics[width=0.95\textwidth]{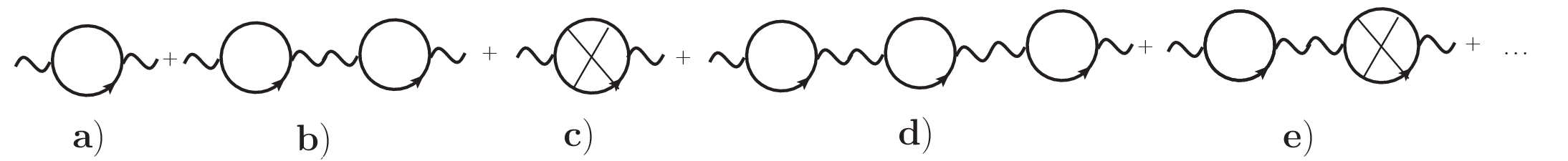}
\caption{The possible combinations of the irreducible diagrams
contributing to up to the sixth order vacuum polarization operator.
Diagrams a), b) and d) are  purely bubble-type diagrams, whereas
diagrams c) and e) are of the so-called   ``mixed type'',~cf.~Fig.~\ref{Irpolar}. } \label{bubble}
 \end{figure}

\begin{figure}[!h]
\vspace*{4mm}
\includegraphics[width=0.25\textwidth]{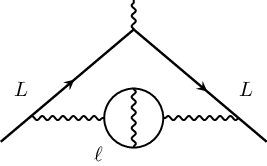} \hspace*{8mm}
\includegraphics[width=0.25\textwidth]{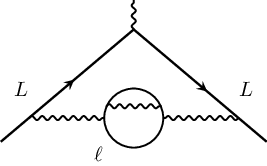} \hspace*{8mm}
\includegraphics[width=0.25\textwidth]{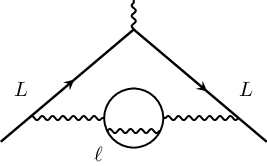}
\caption{ The simplest diagrams with insertions of the vacuum
polarisation  operator of the fourth order with one internal photon
line. The external photon is labeled by     ``$L$'', whereas the
internal lepton in the   loop is denoted as ``$\ell$''.} \label{mixed}
\end{figure}

The quantity  $\widetilde\Pi(k^2)$ in Eq.~(\ref{prop}) includes   all   powers of the irreducible graphs,
     \ba
       \widetilde\Pi(k^2)=    \Pi (k^2)-\Pi^2(k^2)+\Pi^3(k^2)+\cdots , \label{prop1}
       \ea
which  obviously  contains all   orders of the  radiative  corrections. For instance, the powers
of the first diagram in Fig.~\ref{Irpolar} generates a series of     ``bubble-like"\
graphs , labeled as  a), b) and d) in Fig.~\ref{bubble}  (see also
Refs.~\cite{Solovtsova23,Solovtsova-2024}). The  diagram with the crossed loops induces the
so-called ``mixed-bubble" \ types, cf.   diagrams c) and e), which certainly determine  the sixth
and higher orders of the radiative  corrections. The simplest    diagrams of this type  are
depicted in Fig.~\ref{mixed}.

As demonstrated in  Refs.~\cite{Remiddi-1975,Lautrup:1968,Solovtsova23,Solovtsova-2024}, the vertex diagrams with the insertion of only
vacuum polarization operators can equivalently be represented by a  diagram of the second order with the exchange of one but massive photon.
For this type of diagrams the lepton anomaly  $K_L^{(2)}(t)$ is well known in the literature~\cite{berestetski,brodski} and reads as

\begin{equation} \label{KL}
 K_L^{(2)} (t  )= \frac{\alpha}{\pi}
\int\limits_{0}^{1} d x\;\frac{(1-x)x^2}{x^2+ t(1-x)/m_L^2}\;,
 \end{equation}
where $m_L$ is the mass of the  scrutinized lepton and $\alpha$ is the fine structure constant.  In rest of  this paper the
external lepton and also the loop leptons of the same type as the external one are denoted by the  capital letter $L$. The
loop leptons, different from $L$, are denoted by the lowercase $\ell$.

With Eq.~(\ref{KL}), the anomaly $a_L$ can be written as~\cite{Remiddi-1975,Lautrup:1968,Solovtsova23,Solovtsova-2024}
\ba  && a_L=  \frac1\pi\int \frac{dt}{t}
 \Imm \widetilde\Pi(t)  K_L^{(2)}(t)=   \nonumber  \\ &&
  =  \frac1\pi\int \frac{dt}{t}
 \Imm \widetilde\Pi(t)\frac\alpha\pi\int dx\frac{x^2
 (1-x)}{x^2+(1-x)t/m_L^2}= -\frac\alpha\pi\int\limits_0^1 dx (1-x)\tilde \Pi\left( \frac{-x^2}{1-x} m_L^2\right).
\label{double}
\ea
In obtaining Eq.~(\ref{double}), the dispersion relations for the operator $\Pi(t)$ have been  employed.
In the present paper, we consider the polarization operator consisting
of only two types of internal leptons, $L$ and $\ell$. The diagrams
with three different leptons, $L$, $\ell_1$ and $\ell_2$ will be represented elsewhere. Generally,
the polarization operator (\ref{prop1})  for Feynman diagrams with
insertions of $n=p+j$ closed loops, where $p$  and $j$ denote the
number of lepton   loops of $L$ and $\ell$ types, respectively, can be
presented as  (q.v.~Eq.~(\ref{prop1}))
 \ba && \Pi^{ n }(k^2)=  \sum\limits_{p=0}^n (-1)^{n+1}
C_n^p\left[ \Pi^{(L)}(k^2)\right ]^p \left[ \Pi^{(\ell)}(k^2)\right
]^{j=n-p}\equiv \sum\limits_{p=0}^n F_{(p,j)}\left[
\Pi^{(L)}(k^2)\right ]^p \left[ \Pi^{(\ell)}(k^2)\right ]^{j=n-p},
\nonumber
\\[-0.4cm]
&& \label{pandj} \ea
where $C_n^p$ are the familiar combinatorial coefficients and the
quantity $ F_{(p,j)} =(-1)^{p+j+1} C_{p+j}^p $  has been introduced as
to reconcile our formulae with the commonly adopted notation, e.g., in
Refs.~\cite{Aguilar:2008qj,Solovtsova23,Solovtsova-2024}. Inserting
Eq.~(\ref{pandj}) in to Eq.~(\ref{double}) and applying again the
dispersion relations to  $\Pi^{(\ell)}(q_{eff}^2)/q_{eff}^2$, where
$q_{eff}^2=-\displaystyle\frac{x^2}{1-x} m_L^2$, the
corrections $a_L^{(p,j)}$ to the anomaly $a_L$ from diagrams with
$p+j$ loops read as
 \ba &&
 a_L^{(p,j)}=
 F_{(p,j)}\frac\alpha\pi \int\limits_0^\infty\frac{dt}{t} \int\limits_0^1 dx \frac{x^2 (1-x)}{x^2 +(1-x) t/m_L^2}
\left[ \Pi^{(L)} \left(-\frac{x^2 }{1-x)}m_L^2  \right)\right ]^p
\frac1\pi \; \Imm \left[ \Pi^{(\ell)}(t)\right ]^j.~ \label{secnd}
  \ea
Note the factorization of the integrals over $x$ and $t$ for  particular diagram with $p=0$. In this specific case  the
corrections $a_L^{(p,j)}$ in Eq.~(\ref{secnd}) can  already be calculated numerically.  However, since these diagrams with all
leptons $\ell$ in the loops  different  from the external ones, $\ell\neq L$, are too particular,
a  more general and comprehensive  analysis  of $a_L^{(p,j)}$ is hindered.

To further proceed  with analytical calculations,  we consider  the
integrand in Eq.~(\ref{secnd}) as the subject of the Mellin--Barnes
transform, see e.g. Refs.~\cite{Yao,Mellin-book,Davych,Friot:2005cu},
\ba
\frac{x^2 (1-x)}{x^2 +(1-x) t/m_L^2} = \frac{1}{2\pi
i}\int\limits_{c-i\infty}^{c+i\infty} ds \;  \left( \frac{4
m_{\ell}^2}{t}\right)^s\left( \frac{4m_{\ell}^2}{m_L^2}\right)^{-s}
x^{2s}(1-x)^{1-s} \; \Gamma(s) \Gamma(1-s), \label{m-b}
\ea
where $0\,<\,c<\,1$ determines the strip in the complex  plane of $s$  along which  the integrand~(\ref{m-b}) is
 an analytical function;
   the
factor  $4m_\ell^2$ has been introduced for further convenience. By
virtue of the representation (\ref{m-b}), we can write  $
a_L^{(p,j)}$,  Eq.~(\ref{secnd}), as
\ba
a_L^{(p,j)}&&=\frac{\alpha}{\pi}\frac{1}{2\pi i}F_{(p,j)}
\int\limits_{c-i\infty}^{c+i\infty}  ds \; \left
(\frac{4m_{\ell}^2}{m_L^2}\right)^{-s} \Gamma(s)\Gamma(1-s)\int_0^1 dx
\; x^{2s} (1-x)^{1-s} \times\nonumber  \\ &&
 \left[ \Pi^{(L)} \left(-\frac{x^2 }{1-x)}m_L^2  \right)\right ]^p
\int_0^\infty\frac{dt}{t}\left( \frac{4m_{\ell}^2}{t}\right)^s\left(\frac1\pi
\Imm \left [\Pi^{(\ell)}\left(\frac{4m_{\ell}^2}{t}\right)\right]^j\right).
\label{fin}
\ea
Hence,   the Mellin--Barnes transform made it possible to  present the contribution
to the lepton anomaly   from different types of lepton loops in the following factorized form of two Mellin momenta
\ba &&
a_L^{(p,j)}=\frac{\alpha}{\pi}
\frac{1}{2\pi i}F_{(p,j)}
\int\limits_{c-i\infty}^{c+i\infty} ds \;
 \left( \frac{4m_{\ell}^2}{m_L^2}\right)^{-s}
\Gamma(s)\Gamma(1-s)\; \Omega_p(s)R_j(s), \label{fin1} \ea where
\vspace*{-0.2cm}\noindent \ba &&\Omega_p(s)= \int_0^1 dx \; x^{2s}
(1-x)^{1-s} \left[ \Pi^{(L)} \left( -\frac{x^2 }{1-x}m_L^2  \right)
\right ]^p , \label{Omp}
\\[0.2cm]
&& R_j(s)= \int\limits_{4m_\ell^2}^\infty \frac{dt}{t}\left(
\frac{4m_{\ell}^2}{t}\right)^s\frac1\pi \Imm \left
[\Pi^{(\ell)}\left(\frac{4m_{\ell}^2}{t}\right)\right]^j=
2\int_0^1 \delta d\delta(1-\delta^2)^{s-1} \frac1\pi \Imm \left
[\Pi^{(\ell)}\left(\delta\right)\right]^j, \quad
\label{Rj}  \ea
where  the variable  $\delta=\sqrt{1-4m_\ell^2/t}$.

Equations  (\ref{fin1})-(\ref{Omp}) represent the most general expressions for the lepton anomaly from diagrams with vacuum polarization operators consisting of   lepton loops of two types, $L$ and~$\ell$. Note that  in our previous papers we considered only  irreducible diagrams of  the second order, i.e.   pure closed lepton loops. In that case,  for a better illustration of      the  $2(p+j+1)$ orders of corrections, the factors $ \left(\frac{\alpha}{\pi}\right)^{p}$ and  $ \left(\frac{\alpha}{\pi}\right)^{j} $  were emphasized explicitly in the definitions of   $\Omega_p(s)$ and $ R_j(s)$. Since in the present paper we consider also  one-loop    polarization   operators of the fourth order, the indices $p$ and $j$  no longer illustrate unambiguously the order of corrections. For this reason   we include the factors  $ \left(\frac{\alpha}{\pi}\right)^{p}$ and  $ \left(\frac{\alpha}{\pi}\right)^{j} $ into the definition of the Mellin momenta and,  instead, we indicate  the order of  correction   by  a subscript for the corresponding polarization
operators.

As seen from Eq.~(\ref{Rj}),  the Mellin momentum $R_j(s)$  is manifestly independent of the lepton masses. The momentum $\Omega_p(s)$
depends only on  the polarization operator $\Pi^{(L)}(q^2)$ for which the genuine variable is the dimensionless combination
$\displaystyle\frac{4m_L^2}{q^2}$. It implies that in our case $  \Pi^{(L)} \left( -\frac{x^2 }{1-x} m_L^2  \right)=
 \Pi^{(L)} \left( -\displaystyle\frac{4(1-x)}{x^2}   \right)$, i.e., $\Omega_p(s)$ is also independent of the lepton masses.
Consequently, the only dependence on the  masses in $a_L^{(p,j)}$,
Eq.~(\ref{fin1}), enters  through the ratio $m_\ell/m_L$,  which
justifies the commonly adopted  classification of $a_L^{(p,j})$ as~\cite{Kinoshita:1990wp,Jegerlehner:2017gek}
\begin{equation}\label{aL}
a_{L} =  ~A_{1,L}\left( \frac{m_L}{m_L}\right ) +
~A_{2,L}\left( \frac{m_{\ell}}{m_L}\right )  + A_{3,L}\left(\frac{m_{\ell}}{m_L},\frac{m_{\ell_2}}{m_L}\right ),
\end{equation}
\noindent
where the contribution $A_1$ corresponds to diagrams  for which all the  internal loops are formed by the same type of leptons as  the external one; it also includes   diagrams  with exchanges of only one virtual photon without lepton loops. Clearly,  $A_1$ is universal for all kinds of leptons. The  mass-dependent contributions $A_2$ and $A_3$  include diagrams with leptons $\ell\neq L$.

In turn, each of the terms in the sum (\ref{aL}) can be written in the form of expansion  by the fine structure $\alpha$ as
\ba
&& A_{1,L}(m_L/m_L)= A_{1,L}^{(2)}\left(\frac\alpha\pi \right)^1 + A_{1,L}^{(4)}\left(\frac\alpha\pi \right)^2 +
A_{1,L}^{(6)}\left(\frac\alpha\pi \right)^3 + \cdots , \label{A1}\\[2mm]
&& A_{2,L}\left( r={m_{\ell}}/{m_L} \right) = A_{2,L}^{(4)}(r)\left(\frac\alpha\pi \right)^2
+ A_{2,L}^{(6)}(r)\left(\frac\alpha\pi \right)^3 +
A_{2,L}^{(8)}(r)\left(\frac\alpha\pi \right)^4 + \cdots,\label{A23} \\[2mm]
&& A_{3,L}\left(r_1,r_2\right)=
A_{3,L}^{(6)}(r_1,r_2)\left(\frac\alpha\pi \right)^3 + A_{3,L}^{(8)}(r_1,r_2)\left(\frac\alpha\pi \right)^4 +
A_{3,L}^{(10)}(r_1,r_2)\left(\frac\alpha\pi \right)^5 + \cdots ,~~~ \label{A4}
\ea
where $r=r_1= {m_{\ell_1}}/{m_L},~r_2={m_{\ell_2}}/{m_L}$ and the superscripts of the coefficient in the r.h.s.  indicate the
corresponding order of the radiative corrections.
As mentioned,  the leading order correction  to the lepton anomaly
$a_L$  was
calculated, for the first time,  by J.~Schwinger~\cite{Schwinger1948}. In our notation, this corresponds to the coefficient $A_{1,L}^{(2)}=1/2$.
The next coefficients $A_{1,L}^{(4)}$ and $A_{1,L}^{(6)}$ are also known explicitly~\cite{Jegerlehner:2017gek}.
 Concerning the contributions to  $A_1$ from  bubble diagrams, they are known up to  an impressive high order, up to 13 loops, see
\cite{Samuel-n-bubble-1,Samuel-n-bubble-2}. Recent numerical calculations of $A_{1,L}$ up to $\alpha^5$ order can be found
 in Ref.~\cite{Volkov:2024yzc}.

Here below, we focus on further investigations of the mass-dependent coefficients  $A_{2,L}^{(i)}(r)$ in Eq.~(\ref{A23}) by means of the Mellin-Barnes technique.   Previously, in Refs.~\cite{Solovtsova23,Solovtsova-2024} we considered the radiative corrections from diagrams consisting of only pure closed lepton loops, the so-called ``bubble"\,-like diagrams. Now we  complicated  the task by including in the consideration diagrams with loops crossed by  internal photon lines, hereafter referred to as ``mixed"\ diagrams, see Fig.~\ref{mixed}. Hitherto, the exact analytical expressions for this type of diagrams  were derived  only for  the universal coefficient $A_{1,L}^{(6)}=A_{2,L}^{(6)}(r=1)$,  see~Refs.~\cite{Laporta:2009olb,Remiddi-1969},
\begin{eqnarray}
A_{2,L}^{(6)} (1)&&
=\frac{673}{108}-\frac{41\pi^2}{81}-\frac{7\pi^4}{270}-\frac{4}{9}\;
\pi^2 \ln(2) - \frac{4}{9}\;\pi^2\ln^2(2)+\frac{4}{9}\ln^4(2)
\nonumber \\
&& +\frac{32}{3}\;{\rm
Li}_4\left(\frac{1}{2}\right)+\frac{13\;\zeta(3)}{18}=~0.05287065~ ...
\, , \label{Ainiv}
\end{eqnarray}
where  Li$_4$ is the polylogarithm function of the order $4$, ${\rm
Li}_n \left(z\right)=\sum_{k=1}^\infty \left(z^k/k^n \right) $, and $\zeta(3)$ is the Riemann zeta function,
$\zeta(z)=\sum_{k=1}^\infty \left(1/z^k \right)$.

So far,   due to cumber\-someness of the explicit expressions of mixed   diagrams (more than 150 terms for the graphs of Fig.~\ref{mixed}, cf. Ref.~\cite{Laporta:1993ju}), the mass dependent coefficients $A_2^{(6)}(r)$ were studied in detail    in the asymptotic  limits $r\ll 1$ and $r\rightarrow \infty$~\cite{Laporta:1993ju}. Concerning  the  coefficients $A_3^{(6)}(r_1,r_2)$, i.e., diagrams with   all three  different leptons, they were thoroughly  considered  also only in the  asymptotic  limits in Refs.~\cite{Czarneski,Friot:2005cu,Aguilar:2008qj}; however exact expressions for $A_3^{(6)}(r_1,r_2)$ were obtained within a generalized two-fold Mellin-Barnes representation in Ref.~\cite{Ananthanarayan:2020acj} (qq.v.  also references therein quoted).

\section{The sixth order corrections}

The sixth order corrections are generated by  powers of the irreducible operator $\Pi (k^2)$, which includes   graphs of the second and
fourth   orders, i.e. the graphs ~b) and c) in Fig.~\ref{bubble}. The contributions from the graphs of type b)   were analyzed  in
detail in Ref.~\cite{Solovtsova23}. The sought sixth order corrections to $a_L$   from the mixed diagrams are determined by the
irreducible graph c) which are hereafter  referred  to as $\Pi_{4}(k^2)$; the corresponding diagrams are depicted  in Fig.~\ref{mixed}, for
which  $p=0$ and $j=1$ (see also Refs.~\cite{Lautrup:1968,Remiddi-1973,Solovtsova23}). Correspondingly,
  \begin{equation}
\Omega_0(s)=\int_0^1 dx \; x^{2s} (1-x)^{1-s}=
\frac{\Gamma(2-s) \Gamma(1+2s)}{\Gamma(3+s)},
   \label{Omp0}
   \end{equation}
  \begin{equation}
 R_1^{(4)}(s)=
 2\int_0^1 \delta d\delta (1-\delta^2)^{s-1} \, \frac1\pi
\Imm \left [\Pi^{(\ell)}\left(\delta\right)\right] =\left(\frac\alpha\pi\right)^2
 2\int_0^1 \delta d\delta (1-\delta^2)^{s-1}\rho^{(4)}(\delta) ,
\label{Rj11} \\
\end{equation}
  where, for the sake of brevity, we introduced the notation
\begin{equation}
 \left(\frac\alpha\pi\right)^2\rho^{(4)} \left( \delta\right)=
 \frac1\pi \Imm \left [\Pi^{(\ell)}_4\left(\delta\right)\right].
\label{rho4}
\end{equation}

The fourth order ``mixed"   polarization operator $\Pi^{(\ell)}_4\left(\delta \right)$
is well-known in the literature and can explicitly be found in, e.g. Refs.~\cite{Lautrup:1968,sabry,Laursen:1978nw}
\begin{eqnarray} \label{F4u}
\rho^{(4)}\left( \delta\right)&=& \left[\frac{11}{16}+\frac{11
\delta^2}{24}-\frac{7\delta^4}{48}+\left(\frac{1}{2}+\frac{\delta^2}{3}-
\frac{\delta^4}{6}\right)\ln\left(\frac{(1+\delta)^3}{8\delta^2}\right)\right]
\ln\left(\frac{1+\delta}{1-\delta}\right)
\nonumber\\[0.2cm]
&&
+\;\delta\left[\frac{5}{8}-\frac{3\delta^2}{8}-\left(\frac{1}{2}-
\frac{\delta^2}{6}\right)\ln\left(\frac{64\delta^4}{(1-\delta^2)^3}\right)\right]
\nonumber\\[0.2cm]
&&
+\;2\left(\frac{1}{2}+\frac{\delta^2}{3}-\frac{\delta^4}{6}\right)\left[2\;{\rm
Li}_2\left(\frac{1-\delta}{1+\delta}\right)+{\rm
Li}_2\left(-\frac{1-\delta}{1+\delta}\right) \right]
\Theta(1-\delta)\, ,
 \end{eqnarray}
where  ${\rm Li_2}(x)$   denotes the dilogarithm function
\footnote{ Note that in Refs.~\cite{Lautrup:1968,Laursen:1978nw} the operator $\frac{1}{\pi}\Imm \,  \Pi^{(\ell)}_4\left(t\right)$ is
given in terms of the Spence functions $\Phi(x)$ which  are related to the dilogarithm functions  as
$\Phi(x)=-\displaystyle\frac{\pi^2}{12}- {\rm Li}_2(-x) .$}.

Thus, the anomaly $a_L^{(6)}$ becomes
\ba
 \label{anomaly6} &&
a_L^{(6)}=\left(\frac\alpha\pi\right)
\frac{1}{2\pi i}
\int\limits_{c-i\infty}^{c+i\infty} ds \;
 r^{-2s}
\frac{\Gamma(s)\Gamma(1-s)\Gamma(2-s)\Gamma(1+2s)}{\Gamma(3+s)}\; R_1^{(4)}(s)=\nonumber \\   &&
\left(\frac\alpha\pi\right)^3
\frac{1}{2\pi i}
\int\limits_{c-i\infty}^{c+i\infty} ds \;
 r^{-2s}
\frac{2\sqrt{\pi}(1-s)\Gamma(1-s)\Gamma(\frac12+s) }{ (1+s) (2+s)\sin(\pi s)}
2\int\limits_{0}^{1} \delta d\delta (1-\delta^2)^{s-1} \rho^{(4)}(\delta).
  \ea
 With Eq.~(\ref{anomaly6}) the coefficient $A_{2}^{(6)}(r)$ in Eq.~(\ref{A23}) defining the contribution to the  sixth order
  corrections from the diagrams  in Fig.~\ref{mixed} can be represented in the form
\begin{equation} \label{A26bis}
A_{2,L}^{(6)}(r)= \frac{1}{2\pi i} \int\limits_{c-i\infty}^{c+i\infty}
ds \; r^{-2s} \frac{\sqrt{\pi}(1-s)\Gamma(1-s)\Gamma(\frac12+s) }{
(1+s) (2+s)\sin(\pi s)} 2\int\limits_{0}^{1}  d \delta
\delta\, (1-\delta^2)^{s-1}\rho^{(4)}(\delta),
\end{equation}
which, after carrying out integration over $\delta$,  can be re-written as
  \ba
  \label{A26} A_{2,L}^{(6)}(r) = \frac{1}{2\pi
i}\int\limits_{c-i\infty}^{c+i\infty}ds \; r^{-2s}{\cal F}(s)\;,
\ea
where the integrand ${\cal F}(s)$ explicitly reads as
\begin{eqnarray} \label{F}
&&
~~~{\cal F}(s) = \frac{\pi^2 (1-s)}{\sin^2(\pi s)}
\left\{-\frac{72+408s+852s^2+749s^3+199s^4-72s^5-36s^6}{6s^2(1+s)(2+s)^3(1+2s)^2(3+2s)}\right. \nonumber \\
[0.2cm]
 &&
~~~~~~~~  +\frac{ \; \pi^2}{12}\left(\frac{4}{s}-\frac{1}{2+s}\right)
\frac{\Gamma(\frac{1}{2}+s)}
 {\sqrt{\pi}(1+s)(2+s)\Gamma(s)}+\;\frac{2+3s}{3\;s\;(1+s)(2+s)^2(1+2s)} \nonumber
  \\ [0.2cm]
   &&  ~~~~~~~~
\times
\left[\psi\left( s+\frac{1}{2}\right)-3\psi({s})-2\gamma_E+2\ln(2)\right] +\frac{1}{3(\frac{1}{2}+s)(2+s)}
\; \\
[0.2cm]
 && \left. ~~~~ \times
\left[ \frac{-4 \,} {1+s}\;
{_{3}F}_{2}\left(\frac{1}{2},\frac{1}{2},1;\frac{3}{2},\frac{3}{2}+s;1\right)+
\frac{s}{(\frac{3}{2}+s)(\frac{5}{2}+s)}\;{_{3}F}_{2}\left(\frac{1}{2},\frac{1}{2},1;\frac{3}{2},\frac{7}{2}+s;1\right)\right]
\right\}.~~
 \nonumber
 \end{eqnarray}
    \vskip 0.2cm
    \noindent
Above $\gamma_E=0.5772156649\ldots$  denotes the Euler constant, $\psi(s)$ is the well-known  polygamma function  and
${_p}F{_q}(a_1,a_2,...a_p;b_1,b_2,...b_q;1)$  are the generalized  hypergeometric functions of the argument $x=1$. Recall that    in
Eq.~(\ref{A26}) the integration domain is specified by the strip $0<\ c \ <1$ where  the above hypergeometric  functions have been
defined and are analytical functions. A more meticulous analysis of the    functions
$_3F_2\left(\frac{1}{2},\frac{1}{2},1;\frac{3}{2},\frac{3}{2}+s;1\right)$ and
$~_3F_2\left(\frac{1}{2},\frac{1}{2},1;\frac{3}{2},\frac{7}{2}+s;1\right)$ shows that they are valid  in a larger region, namely in the
complex plane $\Ree\,  s> -1$ and  $ \Ree\, s> -3$, respectively, whereas outside this regions these functions are not defined at all.
This can  impede further evaluations of the integrals by the Cauchy residue theorem  in the left semi-plane of  $s$. Therefore, an
analytical continuation of  these functions to the left  semi-plane is necessary.

\subsection{Direct numerical calculations}\label{direct}

Before  proceeding with analytical calculations of the   integral  (\ref{A26}) with the integrand (\ref{F}) by the Cauchy  residue
theorem, we recall that for this particular case considered in the present paper, i.e. for $p=0$ and $j=1$, the coefficient
$A_{2,L}^{(6)}(r)$ can be calculated numerically directly from the definition (\ref{secnd}). In this case, the $x$-direction integration in
Eq.~(\ref{KL}) or   Eq.~(\ref{secnd})   can be carried out analytically providing for the coefficient $A_2^{(6)}(r)$ the   following
expression
 \begin{equation} \label{A3u}
 A_{2,L}^{(6)}(r) = \int\limits_{0}^{1}\frac{2\delta}{1-\delta^2}\;{\widetilde{K}}_2
(\delta,r)\;\rho^{(4)}(\delta)\;d\delta,
 \end{equation}
where   $\rho^{(4)}(\delta)$  is defined by Eq.~(\ref{F4u}) and
${\widetilde{K}}_2 (\delta,r)$ is the result of integration over $x$ in
Eq.~(\ref{KL}) with $t/m_L^2=4r^2/(1-\delta^2)$  (see also  Refs.~\cite{Lautrup:1968,brodski}),

\begin{eqnarray} \label{K2u}
{\widetilde{K}}_2(\delta,r) &=& \frac{1}{2} \left[
1-\frac{8r^2}{1-\delta^2}-\frac{8r^2}{1-\delta^2} \left(1-\frac{2r^2}{1-\delta^2}\right)
\ln \left( \frac{4r^2}{1-\delta^2} \right) \right]
\nonumber\\[0.2cm]
&& + \frac{1}{\sqrt{1-{\displaystyle{\frac{1-\delta^2}{r^2}}}}}
\left[1-\frac{8r^2}{1-\delta^2}+\frac{8r^4}{(1-\delta^2)^2}\right]\ln\left(\frac{1-{\displaystyle\sqrt{1-\frac{1-\delta^2}{r^2}}}}
  {1+\sqrt{1-{\displaystyle\frac{1-\delta^2}{r^2}}}}\right) .
 \end{eqnarray}

In principle, Eqs.~(\ref{A3u}) and  (\ref{K2u}) are already  suitable for numerical calculation of $A_2^{(6)}(r)$. However, these
expressions are not sufficiently convenient as, they do not allow further   refinement investigations of the analytical properties of
$A_{2,L}^{(6)}(r)$, such as the asymptotic behaviour as $r\to\infty$ or $r\ll 1$, dependence on the types of leptons in the loops, comparisons with
results already reported in the literature,  etc. Such kind of analysis  can be performed  only if one has explicit analytical
expressions of  $A_{2,L}^{(6)}(r)$. Nevertheless, Eq.~(\ref{A3u}) can serve as   an extremely   useful tool for crosschecking the numerical
results  obtained from the exact, but rather cumbersome  analytical  expressions  (see below).

Besides this possibility of numerical checks  of  the
exact formulae, another way of testing the results is to  compare  the
analytical expressions with those known  in the literature. Later on we  compute  (numerically) the limit $r\to  1$  and compare  with the
explicit  expression for $A_{2,L}^{(6)} (r=1)$, Eq.~(\ref{Ainiv}). Also, the limits $r\gg 1$ and $r\ll 1$ are employed for comparisons with
the well-known results \cite{Laporta:1993ju}.

\subsection{Analytical calculations: right semi-plane $r>1$}\label{right}


Having computed explicitly the integrand~(\ref{F}), the Mellin integral   (\ref{A26})   can be  calculated straightforwardly by means of
the  Cauchy residue theorem closing the integration contour consecutively  to the right ($r>1$) and  to the left ($r<1$)
semi-planes of the complex variable~$s$. From Eq.~(\ref{F}) one infers that ${\cal F}(s) $ is a singular function in both semi-planes
of~$s$. So, in the right semi-plane ($r >1$) the integrand  ${\cal F}(s)$ possesses  a  pole of the first order at $s=1$ and
multiple poles of the
second order for other positive integers $s$, $s=2,~3,~...~n~...\,$.
The residue at $s=1$ is easily calculable
\begin{equation}
\Res\left[{ r^{-2s} \cal F}(s)\right]_{s=1}=-\frac{41}{486}\;\frac{1}{r^2}\, \label{pole1}.
\end{equation}
The remaining poles, all of the second order,  are located at integer   $s>1$; hence
$A_{2}^{(6)}$ can be written as
\begin{equation}
A_{2,L}^{(6)}(r>1)=\frac{41}{486}\;\frac{1}{r^2}-\sum_{n=2}^{\infty}\Res\left[{r^{-2s} \cal F}(s)\right]_{s=n}.
\label{poles}
\end{equation}
The minus sign in (\ref{poles}) originates from the fact that the integration contour is clockwise. The   residues in $s=2,3\;...$ in
Eq.~(\ref{poles}) have been calculated  by means of the available computer packages,
Wolfram Mathematica 13 and/or Maple 2023, with ability of  symbol manipulation. The result is
\begin{equation}
A_{2,L}^{(6)}(r>1)= C_1(r)+
C_2(r)\ln(r)+\bigg(\frac{7}{2}-8r^2-\frac{115}{18}r^4\bigg)\ln^2(r)+
\Sigma(r)  \, , \label{FinalB}
\end{equation}
where
\begin{eqnarray} \label{C1R}
&&C_1 (r) =\frac{193}{1215r^2}+ \frac{919}{144}-\frac{33}{4}\,
r^2+r\left(\frac{8}{9} +10\;r^2 \right) \left[{\rm
Li}_2\left(\frac{1-r}{1+r}\right)-{\rm
Li}_2\left(-\frac{1-r}{1+r}\right)\right]\nonumber \\[0.2cm]
&& -\frac{16}{9}r\left[{\rm Li}_3\left(\frac{1}{r}\right)-{\rm
Li}_3\left(-\frac{1}{r}\right)\right]+
\left(\frac{7}{4}-4r^2-\frac{115}{36}r^4\right)\bigg[2{\rm
Li}_2\left(\frac{1-\frac{1}{r}}{1+\frac{1}{r}}\right)-2{\rm
Li}_2\left(-\frac{1-\frac{1}{r}}{1+\frac{1}{r}}\right)\nonumber \\[0.2cm]
&&+4\;{\rm Li}_2(1-r)\bigg]
-\left(1+\frac{4}{3}r^2+\frac{23}{9}r^4\right){\rm
Li}_3\left(\frac{1}{r^2}\right)-6 \; r^4{\rm
Li}_4\left(\frac{1}{r^2}\right) -\frac{\pi^2}{3}\bigg\{
\frac{~11}{~144 r^2}+
\frac{7}{8}-\frac{2}{3} r
\nonumber \\[0.2cm]
&& -3r^2 -\frac{15}{2}r^3 -\frac{5869}{72}r^4+
\frac{5984\;r^4-19232\;r^2+20388}{72
\left({\displaystyle1-\frac{1}{r^2}}\right)^{{5}/{2}}}-
\;\frac{6854r^2+367}{72\;r^4
\left({\displaystyle1-\frac{1}{r^2}}\right)^{{5}/{2}}}
 \nonumber \\
&& - 26 r^2\left( 1+\frac{230}{39}r^2 \right) \ln
\left[\frac{1}{2}\left(1+\sqrt{1-\frac{1}{r^2}}\right)\right] +\;
26r^2{_4}F{_3}\left(-\frac{1}{2},1,1,1;2,2,2;\frac{1}{r^2}\right)\nonumber \\
&& -\frac{1}{54
r^2}{_4}F{_3}\left(1,1,\frac{3}{2},3;4,4,4;\frac{1}{r^2}\right)
 + \; \frac{1}{9
r^2}\;{_6}F{_5}\left(\frac{3}{2},3,3,3,3,3;1,1,4,4,4;\frac{1}{r^2}\right)
\! \bigg\},
\end{eqnarray}
\begin{eqnarray} \label{D1R}
&& C_2 (r) =7-\frac{127}{18}r^2-\frac{8}{9}r \bigg[ {\rm
Li}_2\left(\frac{1}{r}\right)-{\rm
Li}_2\left(-\frac{1}{r}\right)\bigg]-\left(1+\frac{4}{3}r^2+\frac{23}{9}r^4\right)
{\rm Li}_2\left(\frac{1}{r^2}\right) \quad
\nonumber\\[0.2cm]
&& ~~~~~  -4r^4 {\rm Li}_3\left(\frac{1}{r^2}\right) -
\frac{\pi^2}{3}\Bigg\{1-10r^2+\frac{206}{3}r^4-
\frac{53-259r^2+206r^4}{3\left({\displaystyle1-\frac{1}{r^2}}\right)^{{1}/{2}}}
  \nonumber\\
&& ~~~~~~ +\; 48r^4 \ln
\left[\frac{1}{2}\left(1+\sqrt{1-\frac{1}{r^2}}\right)\right]
 -30r^2\;{_3}F{_2}\left(-\frac{1}{2},1,1;2,2;\frac{1}{r^2}\right) \Bigg\},
\end{eqnarray}
\begin{eqnarray}
&& \Sigma(r) =\frac{8}{3}\; \sum\limits_{n=2}^{\infty} \bigg\{
\left[\frac{4n^3+n^2-14n-9}{(n+1)^2(n+2)(2n+1)}\;
{_3}F{_2}\left(\frac{1}{2},\frac{1}{2},1;\frac{3}{2},\frac{3}{2}+n;1\right)\right.\nonumber\\
[0.2cm]
 &&\left.
-\;\frac{16n^5+28n^4-104n^3-225n^2-60n+30}
{(n+2)(2n+1)(2n+3)^2(2n+5)^2}\;{_3}F{_2}\left(\frac{1}{2},\frac{1}{2},1;\frac{3}{2},\frac{7}{2}+n;1\right)
\right]\nonumber\\
[0.2cm]
 &&\left. +\; \frac{n-1}{n+1}\left[\;{_3}F{_2}\left(\frac{1}{2},\frac{1}{2},1;\frac{3}{2},\frac{3}{2}+n;1\right)-
 \frac{n(n+1)}{(2n+3)(2n+5)}\;{_3}F{_2}\left(\frac{1}{2},\frac{1}{2},1;\frac{3}{2},\frac{7}{2}+n;1\right)\right]\right.\nonumber\\
[0.2cm]
 &&\left. \times \ln(r^2)
-\; \frac{n-1}{n+1} \bigg[\XD\left(\frac{3}{2}+n\right)
-\frac{n\;(n+1)}{(2n+3)(2n+5)}\;\XD\left(\frac{7}{2}+n\right)\bigg]
\nonumber
\right.
\end{eqnarray}
\begin{eqnarray}
&& -\;\frac{ (n-1)(2n+1)(3n+8)\;\Gamma(n+\frac{1}{2})\;
\pi^{{3}/{2}}}{32{\;(n+1)!(n+2)}} \bigg[\psi(n)-\psi\left(n+\frac{1}{2}\right) \bigg]
-\frac{(n-1)(3n+2)}{8 n(n+1)(n+2)}\nonumber\\
[0.2cm]
 &&
\times\;\bigg[3\;\psi^{(1)}({n})-\psi^{(1)}\left(n+\frac{1}{2}\right)
-\bigg(3\psi(n)-\psi\left(n+\frac{1}{2}\right)+2\gamma_E-2\ln(2)\bigg)\ln(r^2)\bigg]
 \nonumber\\
[0.2cm]
 &&
-\;\frac{9n^5+11n^4-17n^3-31n^2-15n-2}{4 n^2(n+1)^2(n+2)^2(2n+1)}
 \bigg\}
\frac{r^{-2n}}{(n+2)(2n+1)} \, .
\label{sum}
\end{eqnarray}
\vskip 1mm

The quantities $\XD\left(n+\frac32 \right )$ and $\XD\left(n+\frac72 \right )$ in Eq.~(\ref{sum})  are the shorthand notation
for the first derivatives of the hypergeometric function,
\begin{eqnarray} \label{X}
&& \XD\left(\frac{3}{2}+n\right) \equiv \frac{\partial \, }{\partial{s}}\;
 \bigg[{_3}F{_2}\left(\frac{1}{2},\frac{1}{2},1;\frac{3}{2},\frac{3}{2}+s;1\right)
 \bigg]_{s=n} \nonumber \\[0.2cm]
&& ~~~~~~~~~~~ ~~
=\sum\limits_{k=0}^{\infty}
\frac{\Gamma\left(\frac{1}{2}+k\right)}{\sqrt{\pi}\;(1+2k)\;}\frac{\Gamma\left(\frac{3}{2}+n\right)}
{\Gamma\left(\frac{3}{2}+k+n\right)}
\bigg[\psi\left(\frac{3}{2}+n\right)-\psi\left(\frac{3}{2}+n+k\right)\bigg]
 \,.
\end{eqnarray}

The   derivatives   $\XD\left(n+\frac72 \right )$ in Eq.~(\ref{sum}) can be  related with $\XD\left(n+\frac32 \right ) $, Eq.~(\ref{X}),
by merely shifting  the argument  by a factor of two, i.e.  $\XD\left(n+\frac72 \right )=\XD\left((n+2)+\frac32 \right )$. Explicitly,
\,  $\XD\left(n+\frac32 \right ) $  can be   calculated from the integral representation of  the corresponding hypergeometric function
\begin{equation}
 {_3}F{_2}\left(\frac12,\frac12,1;\frac32,s+\frac32;1\right)=
  -\frac{1}{\sqrt{\pi}}  \displaystyle\frac{\Gamma\left(s+\frac32\right)}{\Gamma\left(s+1\right)}\;{ \cal  I}(s),
\label{resder}
\end{equation}
where
\begin{equation}
{ \cal I}(s)\equiv\int\limits_0^1\frac{dy}{y}  (1-y^2)^{s}\ln\bigg[\displaystyle\frac{1-y}{1+y}\bigg].
\label{Iz}
\end{equation}

Then the derivative of Eq.~(\ref{X}) w.r.t.  $s$ at $s=n$ becomes
\begin{equation}
\XD \left(n+\frac32\right) =  \frac{1}{\sqrt{\pi}}\displaystyle\frac{\Gamma\left(n+\frac32\right)}{\Gamma\left(n+1\right)}
\bigg\{
  {\cal I}(n)\left[\psi\left(n+1\right) -\psi\left(n+\frac32\right)\right]- { \cal I}^\prime(n)\bigg\},
  \label{XX}
  \end{equation}
with
\begin{equation}
 { \cal I}^\prime(n)=\int\limits_0^1 \frac{d y}{y} (1-y^2)^n \ln\left[1-y^2\right]\ln\left[\frac{1-y}{1+y}\right].
\label{der}
\end{equation}

Notice  that  the integral (\ref{der}) can be calculated analytically for any  positive integer $n$ so that one can assert that Eq.~(\ref{FinalB}) is the sought manifestly analytical expression for $A_2^{(6)}(r)$. To our knowledge, in so far  in the literature the radiative corrections for the coefficients $A_2^{(6)}(r)$ from diagrams with loops containing one internal photon line have been reported only as approximate expressions for $r\gg  1$ and $r\ll 1$, cf. Ref.~\cite{Laporta:1993ju}. Evidently,  with Eqs.~(\ref{FinalB})-(\ref{sum})  we are in a position to perform expansions  of $A_2^{(6)}(r)$ about any $r=r_0$  up to any  number of terms, necessary for reconciling with other  approximate expressions, e.g., with those reported in  Ref.~\cite{Laporta:1993ju}. The asymptotic of the special functions entering into Eqs.~(\ref{FinalB})-(\ref{sum}) are well known, except for the sum $\Sigma(r)$ that contains derivatives $ \XD\left(\xi\right)$ of the   hypergeometric functions with respect to their  half-integer  parameter $\xi$,    thus  hampering explicit calculations of the asymptotics. Consequently, calculations of $A_2^{(6)}(r\gg 1 )$ require analytical expressions for  $ \XD\left(\xi\right)$ up to the desired order $n$ in the expansion $r^{-2n}$. Following the previously adopted accuracy in   estimates of $A_2^{(6)}(r)$, we restrict the sum (\ref{sum}) up to terms $r^{-12}$, which correspond to  the maximum values of the argument  of   $ \XD(\xi)$~to~$\displaystyle\frac{19}{2}$. The corresponding expressions were calculated explicitly by Eq.~(\ref{XX}) and collected in  Table~\ref{Table-X}.

\begin{table}[ht]
\renewcommand{\arraystretch}{1.2}
{
   \caption{Derivatives $\XD\left(  \xi \right )$,  Eq.~(\ref{XX}), for half-integer $\xi$. }
\begin{tabular}{|c|l|}
\hline  $\xi$&   $~~~~~~~~~~~~~~~~~~~~~\XD\left(\xi \right)$  \\[0.3cm]
\hline
 $~~\frac72 ~~$& ~$\frac{49}{32}+\frac{47 }{128}\pi ^2 -\frac{15}{32}
\pi ^2 \ln (2) -\frac{105 }{64}\zeta(3) $~\\[0.3cm]
\hline
$~~\frac92 ~~$& ~$\frac{3335}{1728}+\frac{319 }{768}\pi ^2-
\frac{35}{64} \pi ^2 \ln(2) -\frac{245 }{128}\zeta(3) $~
 \\[0.3cm]
 \hline
 $~~\frac{11}{2} ~~$& ~$\frac{10423}{4608}+\frac{1879 }{4096}\pi ^2- \frac{315}{512} \pi ^2
 \ln (2)-\frac{2205 }{1024}\zeta(3)$  ~   \\[0.3cm]
\hline
 $~~\frac{13}{2} ~~$& ~$\frac{2940053}{1152000}+ \frac{20417 }{40960}\pi
^2-\frac{693  }{1024}\pi ^2 \ln(2)-\frac{4851 }{2048}\zeta(3)$ ~
\\[0.3cm]
\hline
$~~\frac{15}{2} ~~$& ~$\frac{38888209}{13824000}+ \frac{263111}{491520} \pi
^2 -\frac{3003 }{4096}\pi ^2 \ln(2)-\frac{21021 }{8192} \zeta(3)$~   \\[0.3cm]
\hline
~~$\frac{17}{2} ~~$& ~$\frac{1929621569}{632217600}+ \frac{261395}{458752} \pi^2 -
\frac{6435}{8192}\pi ^2 \ln(2)-\frac{45045}{16384} \zeta(3)$~\\[0.3cm]
 \hline
$\frac{19}{2} ~~$& ~$\frac{33118055953}{10115481600}+ \frac{8842385}{14680064} \pi^2 -
\frac{218790}{262144}\pi ^2 \ln(2)-\frac{765765}{262144} \zeta(3)$~\\[0.3cm]\hline
\end{tabular}
\label{Table-X} }
  \end{table}

With these results, the asymptotic of $A_{2 }^{(6)}(r\gg 1)$,  Eq.~(\ref{FinalB}), reads as follows:
\begin{eqnarray}
&& A_{2,L}^{(6)}
(r\gg 1)=\frac{41}{486}\;\frac{1}{r^2}+\frac{1}{r^4}\left(-\frac{449}{5400}
\ln(r)-\frac{19871}{324000}+\frac{49}{768}
\;\zeta(3)\right) \nonumber \\[0.2cm]
 && ~~~~~~~~~~~ ~~
 +\;\frac{1}{r^6}\left(-\frac{62479}{661500} \ln(r)-\frac{665873}{8890560}+\frac{119}{1920}\;\zeta(3)\right)\nonumber \\ [0.3cm]
 && ~~~~~~~~~~~ ~~
 +\frac{1}{r^8}\left(\;\frac{25993}{291600}\ln(r)-\frac{19963}{293932800}
-\frac{245}{4608 }\;\zeta(3)\right)
 + {\cal {O}}\left(\frac{1}{r^{10}}\right),
  \label{1}
\end{eqnarray}
which perfectly  agrees   with the expansion   earlier  reported in Ref.~\cite{Laporta:1993ju}.

\subsection{Analytical calculations: left semi-plane $r<1$}\label{left}

As seen from Eq.~(\ref{F}), in the left semi-plane the integrand
${\cal F}(s)$ possesses poles at negative integers    $s=0,-1,-2,...,
-n \;...\,$ and negative half-integers
$s=-\frac{1}{2},-\frac{3}{2},-\frac{5}{2}$. The residues in the poles
$s=0$ and $s=-1/2$ can be  easily calculated
\begin{eqnarray} &&
\Res\left[ r^{-2s} {\cal F}(s)\right]_{s=0}  =\frac{1}{4}\ln(r)+\frac{1}{2}\;\zeta(3)-\frac{5}{12} \, ,\label{res0} \\[2mm]
&& \Res\left[  r^{-2s}{ \cal F}(s)\right]_{s=-1/2} =
\left(\frac{79}{27}\pi^2-\frac{16}{9}\pi^2\ln(2) -\frac{13}{18}\pi^3
\right)r.   \label{res12}
\end{eqnarray}
\noindent
As for other poles, due to the ill definition  of the hypergeometric functions $_3F_2\left(\frac{1}{2},\frac{1}{2},1;\frac{3}{2},\frac{3}{2}+s;1\right)$ at  $\Ree\,  s<-1$ and of $~_3F_2\left(\frac{1}{2},\frac{1}{2},1;\frac{3}{2},\frac{7}{2}+s;1\right)$ at $ \Ree \, s<-3$, the  calculations of residues are highly hindered. Consequently,    analytical continuations of $_3F_2\left(\frac{1}{2},\frac{1}{2},1;\frac{3}{2},\frac{3}{2}+s;1\right)$ and $~_3F_2\left(\frac{1}{2},\frac{1}{2},1;\frac{3}{2},\frac{7}{2}+s;1\right)$ in the left semi-plane ($r<1$) of the complex variable $s$
are~called~for.

 \subsubsection{ \bf Analytical continuation of the hypergeometric functions. }

To proceed with the analytical continuation,  let us recall that   the
hypergeometric functions in Eq.~(\ref{resder}) emerged  as   a
result of the integration (\ref{A26})-(\ref{F}) and are defined  by the
integral representations ~(\ref{resder}) and (\ref{Iz}). To extend the
definitions of the corresponding hypergeometric functions in the
regions $\Ree~s <-1$ and  $\Ree~s< -3$, respectively, the integrand
(\ref{Iz}) is  re-written  as
\begin{equation}
(1-y^2)^s\frac{1}{y}\ln\left(\frac{1-y}{1+y}\right)=(1-y^2)^{s-1}
\frac{1}{y}\ln\left(\frac{1-y}{1+y}\right)
-y(1-y^2)^{s-1}\ln\left(\frac{1-y}{1+y}\right),
\end{equation}
which, after integration by parts, allows one to represent the integral $
{\cal I} (s)$, Eq.~ (\ref{Iz}),  as a solution of the following
functional equation
   \begin{equation}
\label{Eq-Iz} {\cal I} (s)={\cal I} (s-1)+
\frac{\sqrt{\pi}}{2s}\frac{\Gamma(s)}
{\Gamma\left(\frac{1}{2}+s\right)}   \Bigg|_{s=n}
\end{equation}
with the boundary condition ${\cal I} (0)=-\frac{\pi^2}{4}$.  The
solution of (\ref{Eq-Iz}) is

\begin{equation}
{\cal I} (n)={\cal I} (0)+\sum\limits_{k=1}^{n}
\frac{\sqrt{\pi}}{2k}\frac{\Gamma(k)}
{\Gamma\left(\frac{1}{2}+k\right)}\; =
  -\frac{\sqrt{\pi}\Gamma(n+1){_3}F{_2}
\left(1,1+n,1+n;\frac{3}{2}+n,2+n;1\right)}{2(1+n)\Gamma\left(\frac{3}{2}+n\right)}\;.
\end{equation}

Changing  the   variable from  $n$  to $s$ and comparing with
Eq.~(\ref{resder}), we get the desired analytical continuation of
${_3}F{_2}\left(\frac{1}{2},\frac{1}{2},1;\frac{3}{2},\frac{3}{2}+s;1\right)$
in the whole semi-plane $\Ree~s<-1$:
\begin{equation}
{_3}F{_2}\left(\frac{1}{2},\frac{1}{2},1;\frac{3}{2},\frac{3}{2}+s;1\right)
\Bigg|_{\Ree~s<-1}
\Longrightarrow\frac{{_3}F{_2}\left(1,1+s,1+s;\frac{3}{2}+s,2+s;1\right)}{2(s+1)}\;.
\label{3F2-new-3}
\end{equation}

The analytical continuation of
${_3}F{_2}\left(\frac{1}{2},\frac{1}{2},1;\frac{3}{2},\frac{7}{2}+s;1\right)$
 immediately follows from  Eq.~(\ref{3F2-new-3}) mentioning that
${_3}F{_2}\left(\frac{1}{2},\frac{1}{2},1;\frac{3}{2},\frac{7}{2}+s;1\right)=
{_3}F{_2}\left(\frac{1}{2},\frac{1}{2},1;\frac{3}{2},\frac{3}{2}+\left(
2+s\right);1\right)$, i.e.

\begin{equation}
{_3}F{_2}\left(\frac{1}{2},\frac{1}{2},1;\frac{3}{2},\frac{7}{2}+s;1\right)
\Bigg|_{\Ree~s<-3} \Longrightarrow \frac{1}{2(s+3)}
{_3}F{_2}\left(1,3+s,3+s;s+4,\frac{7}{2}+s;1\right)\;.\label{3F2-new-7}
\end{equation}

Note that Eqs.~(\ref{3F2-new-3})  and ~ (\ref{3F2-new-7}) are fulfilled as exact
identities for $ \Ree \ s >-1$ and~$ \Ree \ s >-3$,
respectively.

The last effort in  preparing  Eq.~(\ref{F}) for the Cauchy integration in the left semi-plane
is to express the corresponding hypergeometric functions  (\ref{3F2-new-3}) and
(\ref{3F2-new-7}) via the Pochhammer symbols $(a)_k$, viz.
\begin{equation}
{_3}F{_2}\left(1,1+s,1+s;2+s,\frac{3}{2}+s;1\right)=
\sum\limits_{k=0}^{\infty}\frac{(s+1)_k}{(s+\frac{3}{2})_k}\;\frac{s+1}{k+s+1}\,
,
\end{equation}
where $(a)_k$, also known  as
the rising factorials,    are defined as $(a)_k=\Gamma(a+k)/\Gamma(a)$.
Then  the integrand ${\cal F}(s)$  in Eq.~(\ref{A26})
takes the form
\begin{eqnarray} \label{F-left}
&&
 {\cal F} (\Ree \ s <1 ) = \frac{\pi^2 (1-s)}{\sin^2(\pi s)}\;
\left\{-\frac{72+408s+852s^2+749s^3+199s^4-72s^5-36s^6}{6s^2(1+s)(2+s)^3(1+2s)^2(3+2s)}\right. \nonumber \\
[0.2cm]
 &&    +\;\frac{2+3s}{3\;s\;(1+s)(2+s)^2(1+2s)}
\left[\psi\left( s+\frac{1}{2}\right)-3\psi({s})-2\gamma_E+2\ln(2)\right]
\; \nonumber \\
[0.2cm]
 &&
  +\frac{ \; \pi^2}{12}\left(\frac{4}{s}-\frac{1}{2+s}\right)
\frac{\Gamma(\frac{1}{2}+s)}
 {\sqrt{\pi}(1+s)(2+s)\Gamma(s)} +\frac{1}{3(\frac{1}{2}+s)(2+s)}\times
  \\ [0.2cm]
 &&   \left[ -\frac{2}{  (s+1)}
\sum\limits_{k=0}^{\infty}\frac{(s+1)_k}{(s+\frac{3}{2})_k}\;\frac{1}{k+s+1}
 \left.
+\; \frac{s}{2 (\frac{3}{2}+s) (\frac{5}{2}+s)}
 \sum\limits_{k=0}^{\infty}\frac{(s+3)_k}{(s+\frac{7}{2})_k}\;\frac{1}{k+s+3} \right ]
\right\}.~~
 \nonumber
 \end{eqnarray}
    \vskip 0.2cm
\noindent Observe that the Euler gamma functions in the definition  of the Pochhammer symbols induce  additional singularities  in the left semi-plane in Eq.~(\ref{F-left}). In turn, in   counting the residues,   these singularities give rise to  double sums over $n$ and $k$ in the final integration. All-together, the integrand in the left semi-plane is singular  with poles of different orders at negative integers $s=0, -1,-2,-3,\ldots$  and  negative half-integers $s=-1/2,-3/2,-5/2\ldots \;$. Then $A_{2,L}^{(6)} (r)$  acquires the form
\begin{equation} \label{Res-left}
A_{2,L}^{(6)} (r)=\sum\limits_{n=0}^{\infty}\Res\left[{ r^{-2s} \cal
F}(s)\right]_{s=-n}+\sum\limits_{n=0}^{\infty}\Res\left[{ r^{-2s} \cal
F}(s)\right]_{s=-(2n+1)/2} \, ,
\end{equation}
where the residues at $n=0$ and $n=-1/2$   were already listed in
Eqs.~(\ref{res0}) and (\ref{res12}), respectively. Other residues were calculated by means of the package Wolfram Mathematica~10.
Explicitly the result is
\begin{eqnarray}
A_{2,L}^{(6)}(r<1)&=&D_{1}(r)+D_{2}(r)\ln(r^2)+D_{3}(r)\ln^2(r)+4r^4\ln^3(r)
+\sum\limits_{n=3}^{\infty}d_{1}(n,r)r^{2n}\,  \nonumber \\
&+&  \sum\limits_{n=3}^{\infty}d_{2}(n,r)r^{2n}\, + \sum\limits_{n=3}^{\infty}\sum\limits_{k=0}^{n-2}d_{3}(n,k,r)r^{2n}\,
+\sum\limits_{n=4}^{\infty}\sum\limits_{k=0}^{n-4}d_{4}(n,k,r)r^{2n} \; ,
 \label{AB}
\end{eqnarray}
where
\begin{eqnarray}
&&D_{1}(r) = -\frac{5}{12}+\frac{179 r^2}{12}-\frac{38395
r^4}{3888}+ \aE \left(\frac{13 r^2}{9}+\frac{469
r^4}{324}\right) +\frac{4}{9} r^4 \ln^4(2)+\frac{32}{3} r^4 {\rm
Li}_4\left(\frac{1}{2}\right)
\nonumber \\
&&
-\frac{5 \pi ^4 }{54}r^4~  + \frac{
8r}{9}\left(1+\aE +\frac{45}{4} r^2\right)\left[{\rm
Li}_2 \left(\frac{1-r}{1+r}\right)-{\rm
Li}_2\left(-\frac{1-r}{1+r}\right)\right]-g(r){\rm Li}_2\left(1-r^2\right)
\nonumber \\
&& ~+\frac{8}{9}\; r \big[{\rm
Li}_3(r)-{\rm Li}_3(-r)\big]+\left(-\frac{3}{2}+\frac{2
r^2}{3}-\frac{85 r^4}{18}-\frac{8 \aE r^4}{3}\right){\rm Li}_3(r^2)+ 12 r^4 {\rm Li}_4\left(r^2\right)
\nonumber \\
&&
~ +\left(\frac{1}{2}-9 r^2+15 r^4\right)
\zeta(3)+ \frac{\pi ^2 }{3}\left\{-\frac{7}{8}+\frac{64 r}{9}
 -\frac{25 r^2}{3} -\frac{119 r^3}{6}+\frac{101r^4}{8}
\right.
\nonumber \\
 &&
 \left.
~ + \;\frac{40r^2\left(3-{10 r^2}+{11 r^4}-{4
r^6}\right)}{9\left(1-r^2\right)^2}+\frac{\aE}{6}\left(1-4r-4 r^2+\frac{13
r^4}{3}\right)
\right. \nonumber \\
 &&
 \left.
-4 r^4 \big[{\rm Li}_2(r)-{\rm Li}_2(-r)\big] +\left(1-4
r^2+\frac{13 r^4}{3}\right){\rm arctanh} (r)+ \frac{16}{15} r^6
\left[2{\;_ 3}F{_2}\left(1,1,1;2,\frac{7}{2};r^2\right)
\right.\right. \nonumber \\
 &&
 \left. \left.
- \;{_3}F{_2}\left(1,1,1;3,\frac{7}{2};r^2\right) -{_
3}F{_2}\left(1,1,2;3,\frac{7}{2};r^2\right)\right] -\frac{16}{3} r
\ln(2)
 -\frac{4}{3} r^4\ln^2(2)
\right.
 \nonumber \\
 &&
  \left.
~ +\frac{2}{3} r \ln(1-r^2)
 \right\}+
\pi ^3 \left\{\sqrt{1-r^2}\;
r\left(-\frac{13}{18}+\frac{53}{9}r^2\right) -\frac{5}{3}
\left(2+\frac{1}{2}r^2 \right)r^3
\right. \nonumber \\
 &&
 \left.
~ +\frac{19}{3} r^4
\arcsin \left(\frac{1-\sqrt{1-r^2}}{2}\right)^{1/2}
-\frac{11}{6} r^4 \arcsin(r)+
\frac{5}{3}r^4\left(\frac{1-\sqrt{1-r^2}}{1+\sqrt{1-r^2}}\right)^{1/2}
\right.
 \nonumber \\
 &&
\left.~~ \times
\left(\frac{5}{2}+\frac{\sqrt{1-r^2}}{2}\right)\right\},
\end{eqnarray}
\noindent
\begin{eqnarray}
&&D_{2}(r) =-\frac{1}{4}-\frac{4 \pi^2}{9}r-\frac{13
}{18}r^2-\frac{\aE }{9} r^2 \left(10+ \frac{79}{3} r^2\right)
+\frac{2425 }{162}r^4+2 \pi ^2 r^4 -4 r^4 \zeta(3) + \frac{16}{9} r
\nonumber \\
 &&
\times \left[{\rm
Li}_2 \left(\frac{1-r}{1+r}\right)
 -{\rm
Li}_2\left(-\frac{1-r}{1+r}\right)\right]
+2 \left(1-\frac{4 r^2}{3}+\frac{31
r^4}{9}+\frac{4 \aE }{3}r^4\right)
{\rm Li}_2(r^2)-12 r^4 {\rm Li}_3(r^2)\, ,
\nonumber \\
\end{eqnarray}
\begin{eqnarray}
&&D_{3}(r) =\frac{13 r^2}{3}-\frac{161 r^4}{9}-\frac{8}{9} r ~{\rm
arctanh}(r)+\left(1-4r^2+\frac{13 r^4}{3}\right)
\ln\left(1-r^2\right)+4r^4 {\rm Li}_2(r^2)\, ,
\nonumber \\
\end{eqnarray}
where
 $\aE = \gamma_ {E}- \ln(2)$ and
$g(r) =-\frac{7}{4}+4 r^2+\frac{115 r^4}{36}+\aE
\left(\frac{1}{3}-\frac{4 r^2}{3}+\frac{13 r^4}{9}\right)$.

The coefficients $d_i$ in the sums (\ref{AB}) are

\begin{eqnarray}
&&~~~~ {d}_{1}(n,r) =\frac{(-1)^n}{\sqrt{\pi}}(-3+n) ! \Gamma
\left(\frac{3}{2}-n\right) \frac{(1+n)(-8+3n)}{3(n-2)(2n-1)}
\nonumber \\
&&~~~~~\times\bigg\{-\frac{P_1(n)}{Q_1^{2}(n)}-\frac{2 P_2(n) }{Q_1(n)}-2\ln^2(2)-
\frac{1}{2}\left[\psi\left(\frac{3}{2}-n\right)-\psi(n) \right]^2
\nonumber \\
&&~~~~~-\left(\frac{P_2(n)}{Q_1(n)}+2\ln(2)\right)\left[\psi\left(\frac{3}{2}-n\right)-\psi(n)-\ln(r^2)\right]
\nonumber \\
&&~~~~~
+\left[\psi\left(\frac{3}{2}-n\right)-\psi(n)\right]\ln(r^2)- \frac{1}{2}\left[\psi^{(1)}\left(\frac{3}{2}-n\right)+\psi^{(1)}(n) \right]-\frac{1}{2}\ln^2(r^2)
 \bigg\}\, , \\[0.2cm]
&&{d}_{2}(n,r) =\frac{(1+n)(-2+3n) }{3(n-2)Q_2(n)}
\bigg\{\bigg[\frac
{18\,{n}^{5}-22\,{n}^{4}-34\,{n}^{3}+62\,{n}^{2}-30\,n+4
}{ \left( 1+n \right)  \left( -2+3\,n \right)Q_2(n)} -\ln(r^2) \bigg]
\nonumber \\
&&~~~~~~~~~~~ \times
\left[\psi\left(\frac{1}{2}-n\right)-3 \psi\left(1+n \right)\right]
 +\psi^{(1)}\left(\frac{1}{2}-n\right)+3\psi^{(1)}\left(1+n\right)
 \bigg\} \, ,\\[0.2cm]
%
&&{d}_{3}(n,k,r) =\frac{4}{3}\;\frac{(-1)^k(-1+n)!(1+n)^2
\Gamma\left(\frac{3}{2}-n\right) }{(-1+n-k)!
\Gamma\left(\frac{3}{2}-n+k\right){Q}_{3}(n,k)}\bigg\{\frac{P_{3}(n,k)}{{Q}_{3}(n,k)}
+\ln(r^2)
\nonumber \\
&&~~~~ ~~~~  ~~~~
+\psi\left(\frac{3}{2}-n+k\right)-\psi\left(\frac{3}{2}-n\right)+\psi(n)
-\psi(n-k) \bigg\}\, ,
\\[0.2cm]
&&{d}_{4}(n,k,r) =-\frac{4}{3}\;\frac{(-1)^k(-3+n)!n^2(1+n)^2
\Gamma\left(\frac{7}{2}-n\right)}{(-3+n-k)!
\Gamma\left(\frac{7}{2}-n+k\right)Q_4
(n,k)}\bigg\{\frac{P_{4}(n,k)}{{Q}_{4}(n,k)} +\ln(r^2)
\nonumber \\
&&~~~~~~~~~~~ +
\psi\left(\frac{7}{2}-n+k\right)-\psi\left(\frac{7}{2}-n\right)+\psi(n-2)
-\psi(n-k-2)\bigg\}\, ,
\label{d4}
\end{eqnarray}
where the summations  over $k$ and $n$ in the   double sums, as mentioned,
occurred  due to  the summation over $k$ of the
Pochhammer symbols,   Eq.~(\ref{F-left}). Above, the corresponding polynomials $P_i(n)$  and  $Q_i(n)$ are as follows
 \begin{eqnarray}
&& {P}_{1}(n)=
108\,{n}^{8}-702\,{n}^{7}+453\,{n}^{6}+5492\,{n}^{5}-13232\,{n}^{4}+
6440\,{n}^{3}+9425\,{n}^{2}-11180\,n+3296
,
\nonumber \\
&&{P}_{2}(n)= 12\,{n}^{4}-39\,{n}^{3}-32\,{n}^{2}+143\,n-74,
\nonumber \\
&& P_3(n,k)=  \left( 4\,{n}^{3}-{n}^{2}-14\,n+9 \right) k-6\,{n}^{4}+10\,{n}^{3}+13
\,{n}^{2}-28\,n+11,
\nonumber \\
&& P_4(n,k)=  \left( 16\,{n}^{5}-28\,{n}^{4}-104\,{n}^{3}+225\,{n}^{2}-60\,n-30
 \right) k-24\,{n}^{6}+120\,{n}^{5}-46\,{n}^{4}
\nonumber \\
&& ~~~~~ ~~~~~  ~~~ -548\,{n}^{3}+812\,{n}^
{2}-180\,n-90, \nonumber \\
%
&& {Q}_{1}(n)= (-2+n)(-1+n)(1+n)(-1+2n)(-8+3n),
\nonumber \\
&& {Q}_{2}(n)= (-2+n)(-1+n)n(1+n)(-1+2n)(-2+3n),
\nonumber \\
&& {Q}_{3}(n,k) =(-2+n)(-1+n)(1+n)(-1+2n)
 (-1+n-k),
 \nonumber \\
&& {Q}_{4}(n,k) = \left( -2+n \right)  \left( -5+2\,n \right)  \left( -3+2\,n \right)
 \left( -1+2\,n \right)  \left( -3+n-k \right) n \left( 1+n \right)
.
 \nonumber
\end{eqnarray}

\vskip 0.2cm
As seen, the anomaly  $A_2^{(6)}(r)$ defined by Eqs.~(\ref{AB})-(\ref{d4}) is too  lengthy to be   appropriate  for an explicit   analysis  of its analytical properties. At this stage, we can only compare our results with the approximate results reported early in the literature, e.g., in Ref.~\cite{Laporta:1993ju}.

As mentioned, up to now the sixth order corrections to the anomaly due to insertions of mixed loops were analysed as asymptotic at low, $A_2^{(6)}(r\ll 1)$, and high, $~A_2^{(6)}(r\gg 1)$, values  of   $r$. We can compare our results  with  these  asymptotics, e.g.  $A_2^{(6)}(r\ll 1)$, by  restricting the summation  in Eq.~(\ref{AB}) with only a few first residues,
$s=0,-1,-2,-3,-\frac{1}{2},-\frac{3}{2},-\frac{5}{2}$.

The result is
\begin{eqnarray}
&& A_{2,L}^{(6)} (r\ll 1) =-\frac{1}{4}\ln (r)+\frac{1}{2} \;
\zeta(3)-\frac{5}{12}+\pi^2\left(\frac{79}{27}-\frac{13}{18}\pi-\frac{16}{9}\ln(2)\right)r
\nonumber \\
 &&
+ \left(6\ln^2(r)
+3\ln(r)+\frac{35}{3}+\pi^2-9\zeta(3)\right)r^2+\pi^2\left(\frac{35}{12}\pi-\frac{50}{9}\right)r^{3}\nonumber \\
 &&+\bigg[4\ln^3(r)-\frac{27}{2}\ln^2(r)+\left(\frac{107}{4}+2\pi^2-4\zeta(3)
 \right)\ln(r)-\frac{4}{9}\pi^2\ln^2(2)-\frac{5}{54}\pi^4
\nonumber \\
 &&
 +\frac{32}{3}{\rm
Li}_4\left(\frac{1}{2}\right)+\frac{4}{9}\ln^4(2)+15\zeta(3)-\frac{9}{4}\pi^2-\frac{3739}{144}
  \bigg]r^4-\frac{7}{3}\pi^2\left(\frac{11}{15}-\frac{1}{16}\pi\right)r^{5}
\nonumber \\
 &&+
  \left(\frac{116}{45}\ln^2(r)-\frac{1127}{225}\ln(r)+\frac{42343}{10125}+\frac{58\pi^2}{135}
  \right)r^6+{\cal {O}}(r^{7})\; .
 \label{A3less_Laporta}
\end{eqnarray}
A comparison of  the terms up to $ {\cal {O}}\left(r^{5}\right)$ in  Eq.~(\ref{A3less_Laporta})
with the respective  expression reported in  Ref.~\cite{Laporta:1993ju}
shows that they   exactly reproduce the results of Ref.~\cite{Laporta:1993ju}.  This is an additional
confirmation of the validity of the exact, but lengthy, expressions presented by Eqs.~(\ref{AB})-(\ref{d4}).

\section{Numerical results}\label{num}

Here below we perform some checks of the validity of the obtained exact expressions  by comparing the results from   Eqs.~(\ref{AB})-(\ref{d4}) with the corresponding  direct numerical calculations of the integrals in Eqs.~(\ref{A3u})-(\ref{K2u}). We have made such  checks at several values of $r$ and  obtained perfect agreements  between two methods of calculations.  This  persuades us  of high confidence of the validity in our analytical results. However, a comment is in line here. We found  that  the   sums, Eqs.~(\ref{sum}) and (\ref{AB}),  in the analytical expression are rather poorly convergent. Consequently, for reliable calculations it is required to take into account a large number, several  hundred or even thousand,
of terms in the sums.

For instance,  to obtain the value  $A_{2,L}^{(6)}
({1}/{2})\approx 0.13592598875835838363$ at $r=1/2$ provided by
Eq.~(\ref{A3u}), it is necessary to include $\sim$~50 terms in to
Eqs.~(\ref{AB})-(\ref{d4}).   A slightly
better convergence of sums occurs at larger $r$. Thus, at $r=2$  to assure
the value $A_{2,L}^{(6)}
( {2})\approx  0.017086812024867475167$ it
is sufficient to take into account $\sim 30$ terms. The
situation worsens as $r\to 1$, where to assure  only three
significant digits, in comparison with the calculations  by
Eq.~(\ref{Ainiv}), one already needs ~$\sim 150$ terms, whereas for
a reliable accuracy
 more than one thousand terms are necessary.

\begin{figure}
 \includegraphics[width=0.47\textwidth]{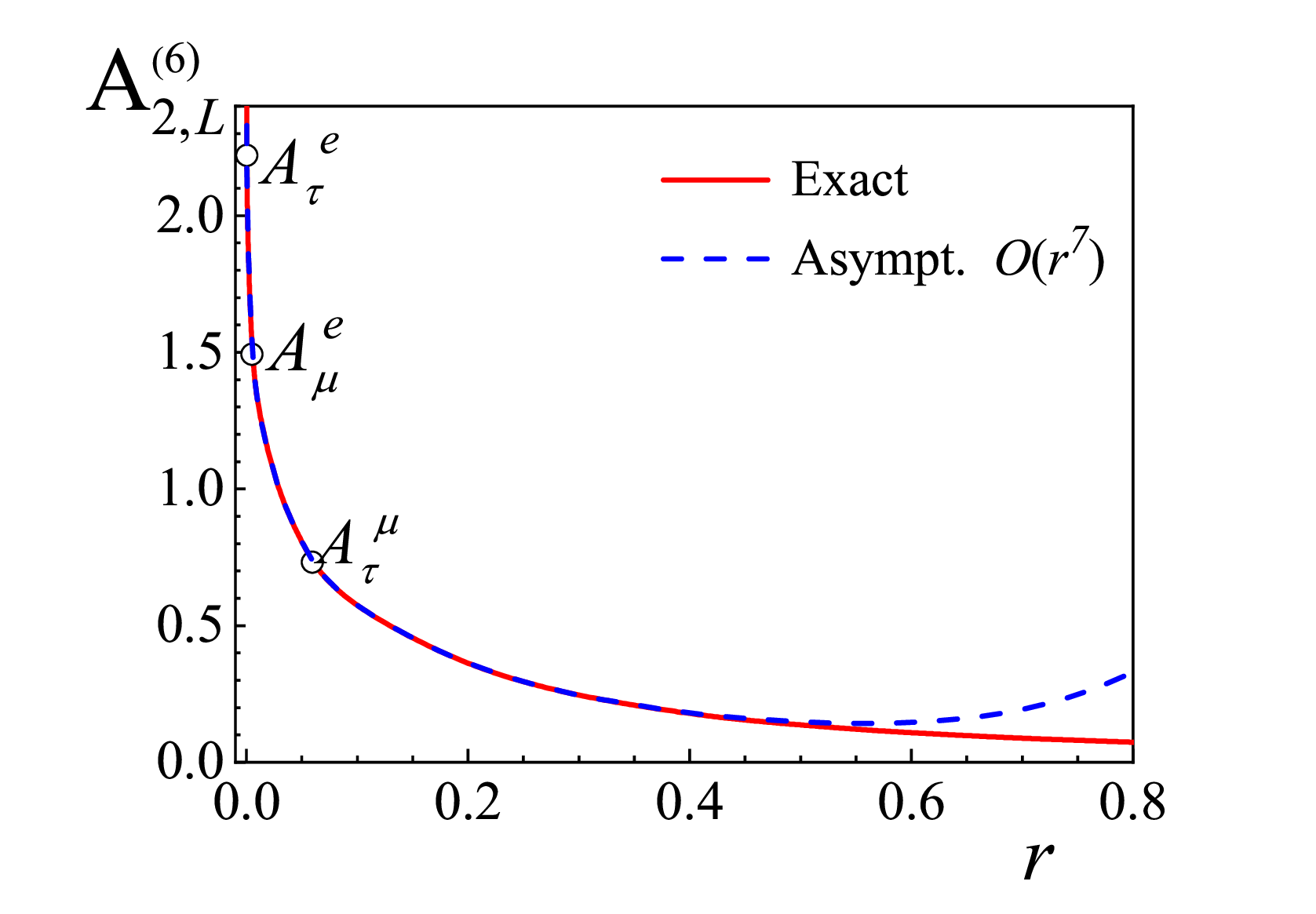}
 \includegraphics[width=0.47\textwidth]{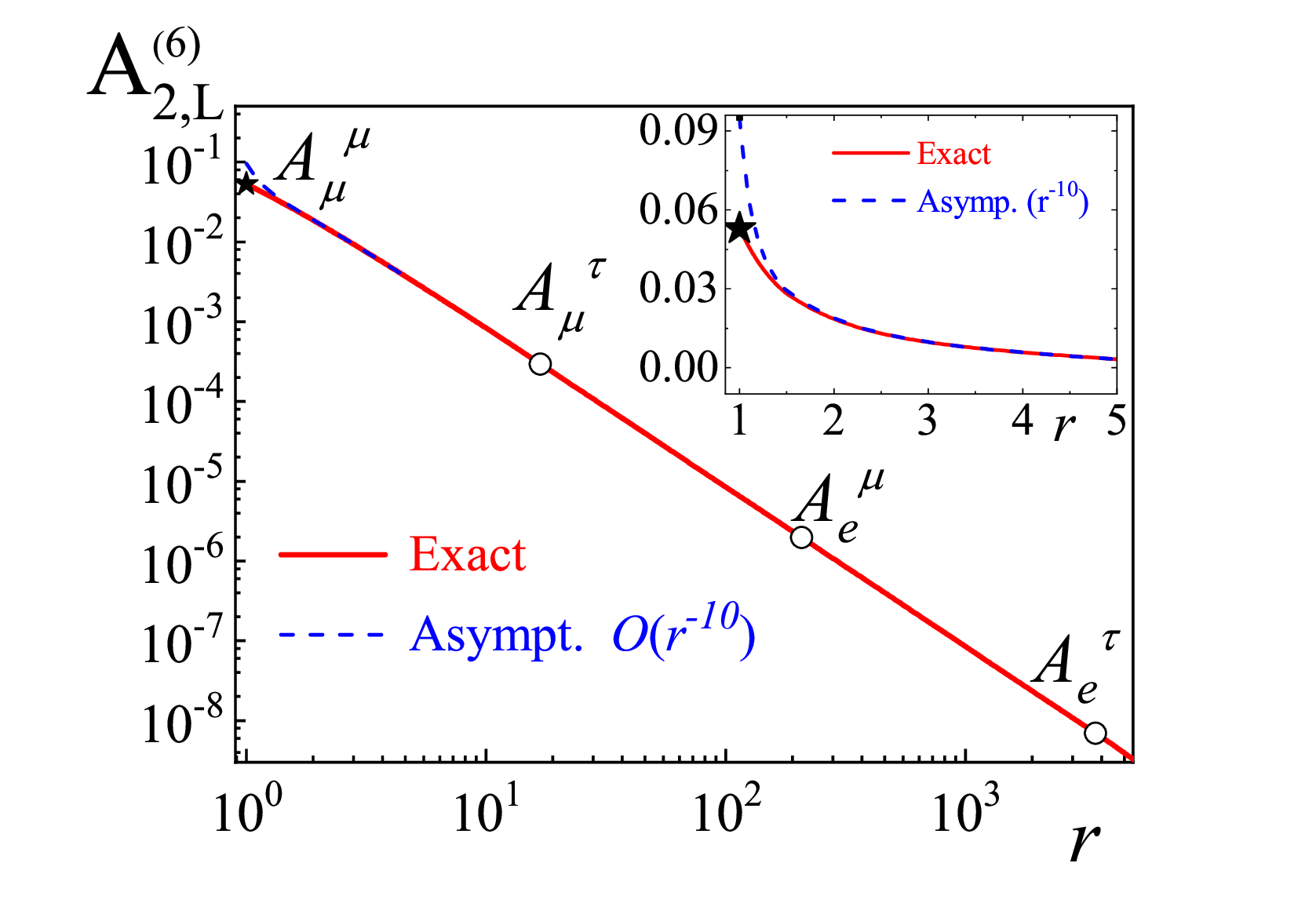}
\caption{(color online)  The sixth order corrections $A_{2,L}^{(6)}(r)$ to the lepton
anomaly due to insertions of the fourth order polarization operator
with one internal photon line. The solid curves correspond to calculations
by exact analytical expressions, the dashed curves are the results of
calculations by asymptotic expansion,  Eq.~(\ref{A3less_Laporta}) for
the left panel and Eq.~(\ref{1}) for the right panel. The open
circles and their  associated  labels $A_L^\ell$ point to
the physical values of $A_{2,L}^{(6)}(r)$ at  the physical ratio
$r=m_\ell/m_L$. The star corresponds to the universal value of
$A_{2,L}^{(6)}(r=1)$.  For a better vision of the behaviour at small $r$,
 the magnified  area $1<r<5$ is highlighted  in the upper corner where the linear scale  for both axes is  employed.
} \label{246012}
 \end{figure}

The behaviour of  $A_{2,L}^{(6)} (r)$  in dependence on $r$ is
presented on Fig.~\ref{246012}. The solid curves reflect the results
of calculations by the exact formulae (\ref{FinalB})-(\ref{sum}),
right panel, and by Eqs.~(\ref{AB})-(\ref{d4}), left panel. The
dashed curves are the results of calculations of the asymptotic  for
$r>1$, Eq.~(\ref{1}), and $r<1$,   Eq.~(\ref{A3less_Laporta}), the
right and left panels, respectively. The open circles and
therein associated labels $A_L^\ell$ point to the physical values
of $A_{2,L}^{(6)}(r)$ at the physical lepton mass ratios $r=m_\ell/m_L$.
  In our actual calculations these   ratios  correspond to those  recommended by the  International
Committee on DATA (CODATA 2018)~\cite{CODATA:2018}, namely:
$ m_e/m_\tau=0.000287585(19)$,  $
m_e/m_\mu=0.00483633169 (11)$, $ m_\mu/m_\tau=0.0594635(40)$,  $
m_\tau/m_\mu=16.8170(11)$, $ m_\mu/m_e=206.7682830(46)$ and $
m_\tau/m_e=3477.23(23)$.
%


From  Fig.~\ref{246012} and from the corresponding numerical analysis, one can conclude that
the exact analytical results  practically coincide with the results by
approximate formulae~(\ref{A3less_Laporta}) in the interval  $0<r<0.2 $ and  by Eq.~(\ref{1})
in the interval  $2<r<\infty $, respectively. More precisely, the validity
 of the  asymptotic expansions and the limits of their applicability
 can be illustrated
if one  defines the relative deviation  $\varepsilon_L(r)$ of the
approximate calculations  from the exact results as
\begin{equation}
\varepsilon_L(r)={\bigg |A_{2,L\ asymp.}^{(6)}(r)  - A_{2,L\ exact}^{(6)}(r)\bigg |}/
{A_{2,L\ exact}^{(6)}(r)}.
\nonumber
\end{equation}
Then  for, e.g.,  the muon anomaly
 the  maximum contribution   is  from the fourth  order electron  vacuum polarization operator.
In this case,  ($L=\mu$ and $\ell=e $)
 $ A_{2,\mu\ exact}^{(6)  }(r_e) \simeq  1.49367182340837205$,
 whereas Eq.~(\ref{A3less_Laporta}), keeping terms up to ${\cal O}(r^4)$
 (this approximation corresponds to the one reported in  Ref.~\cite{Aguilar:2008qj}), results in
$ A_{2,L\ asymp}^{(6) }\big(  \sim{\cal O}(r^4) \big)=1.49367182344.$
In  terms of the relative errors $\varepsilon_\mu(r_e)$ this corresponds to
 $\varepsilon_\mu(r_e) \big(  \sim{\cal O}(r^4) \big) \approx 2.1 \cdot 10^{-11}$. The relative errors rapidly decrease if in Eq.~(\ref{A3less_Laporta}) one keeps terms up to
${\cal O}(r^6)$. In this case
$A_{2,L\ asymp}^{(6)}\big(  \sim{\cal O}(r^6) \big)=1.49367182340837220$ and
  consequently,  $\varepsilon_\mu(r_e)(\sim{\cal O}(r^6) \big )\approx
1.0 \cdot 10^{-16}$. Hence, in calculations of the electron corrections to the muon anomaly it is quite sufficient
to restrict oneself to terms
  $ \sim {\cal O}(r^4)$ which assure  accuracies higher than the experimental  errors
   related to the measured~\cite{CODATA:2018} ratio of electron  to muon masses  $ \Delta r\sim 10^{-10}$.
To estimate how far from $r\to 0$ one can apply the approximate formula, Eq.~(\ref{A3less_Laporta}),
we compare the exact results with the expansions
   $ \sim {\cal O}( r^4)$ and $ \sim{\cal O}(r^6)$ at $r=0.1$. We obtained that keeping
   terms  $\sim  {\cal O}(r^4)$,    Eq.~(\ref{A3less_Laporta})  assures only four significant digits
  while keeping terms   $ \sim{\cal O}(r^6)$, the approximate formula provides much more accurate results, namely up
  to seven significant digits in the interval $(0<\, r\, < 0.1)$. This accuracy is above  the experimental measurements in this  interval. An analogous situation occurs also in the region  $r>2$, cf. Table~\ref{table-1}.
\begin{table}[htbp]
\addtolength{\tabcolsep}{-1pt}
 \caption{ The relative deviation $\varepsilon(r)$
  of the asymptotic results, Eqs.~(\ref{A3less_Laporta}) if $r<1$, and (\ref{1}) if $r>1$, from the exact calculations at the mass ratio
  $r= {m_\ell}/{m_L}$ corresponding to  the really existing
 leptons~\cite{CODATA:2018}.  }
\begin{tabular}{|c|c|c|c||c|c|c|c| c|} \hline
     \multicolumn{4}{| c||}{$r<1$}&\multicolumn{4}{|c|}{$r>1$}\\\hline
 {~mass ratio } & $ m_e/m_\tau $ & $m_e/m_\mu $ & $ m_\mu/m_\tau $ & mass ratio& $m_\tau/m_\mu  $& $ m_\mu/m_e $& $ m_\tau/m_e $\\ \hline
~$r$ & $ \,0.000287585\, $ & $ \, 0.00483633169 \,  $& $\, 0.0594635\,$ &
\multicolumn{1}{|c|}{r  }& $ 16.817  $& $ 206.768283  $& $ 3477.23 $
 \\\hline
 ~$   \varepsilon ({\cal O}(r^4))$ & $1.1\times 10^{-17}$& ~$2.1\times 10^{-11}$ &$1.0\times 10^{-5}$~
  &  {$\varepsilon({\cal O}(r^{-8}))$} &~ $1.4\times 10^{-7}$~ &~$7.3\times 10^{-14}$~&  $4.9\times 10^{-21}$
 \\ \hline
~$   \varepsilon ({\cal O}(r^6))$~ & $1.9\times 10^{-25}$&~$1.1\times 10^{-16} $ ~& $7.3\times 10^{-9}$~
 &  {$\varepsilon({\cal O}(r^{-10}))$} & ~$3.6\times 10^{-8}$~&~ $1.1\times 10^{-14}$~
 &$4.5\times 10^{-22}$
   \\ \hline
\end{tabular}
\label{table-1}
\end{table}

\subsection{bubble-like contribution}\label{bubb}
Now we shall  discuss the contribution of other sixth order corrections to $A_{2,L}^{(6)}(r)$ from insertions of the fourth order
polarization operators. As mentioned, besides the diagrams defined in  Fig.~\ref{mixed}, there is another type of the fourth order
polarization operators, namely, the operator  consisting of two closed lepton loops as depicted in Fig.~\ref{bubble}b, usually referred to
as ``bubble-like'' diagrams.
\begin{figure}
 \includegraphics[width=.65\textwidth, height=.43\textwidth]{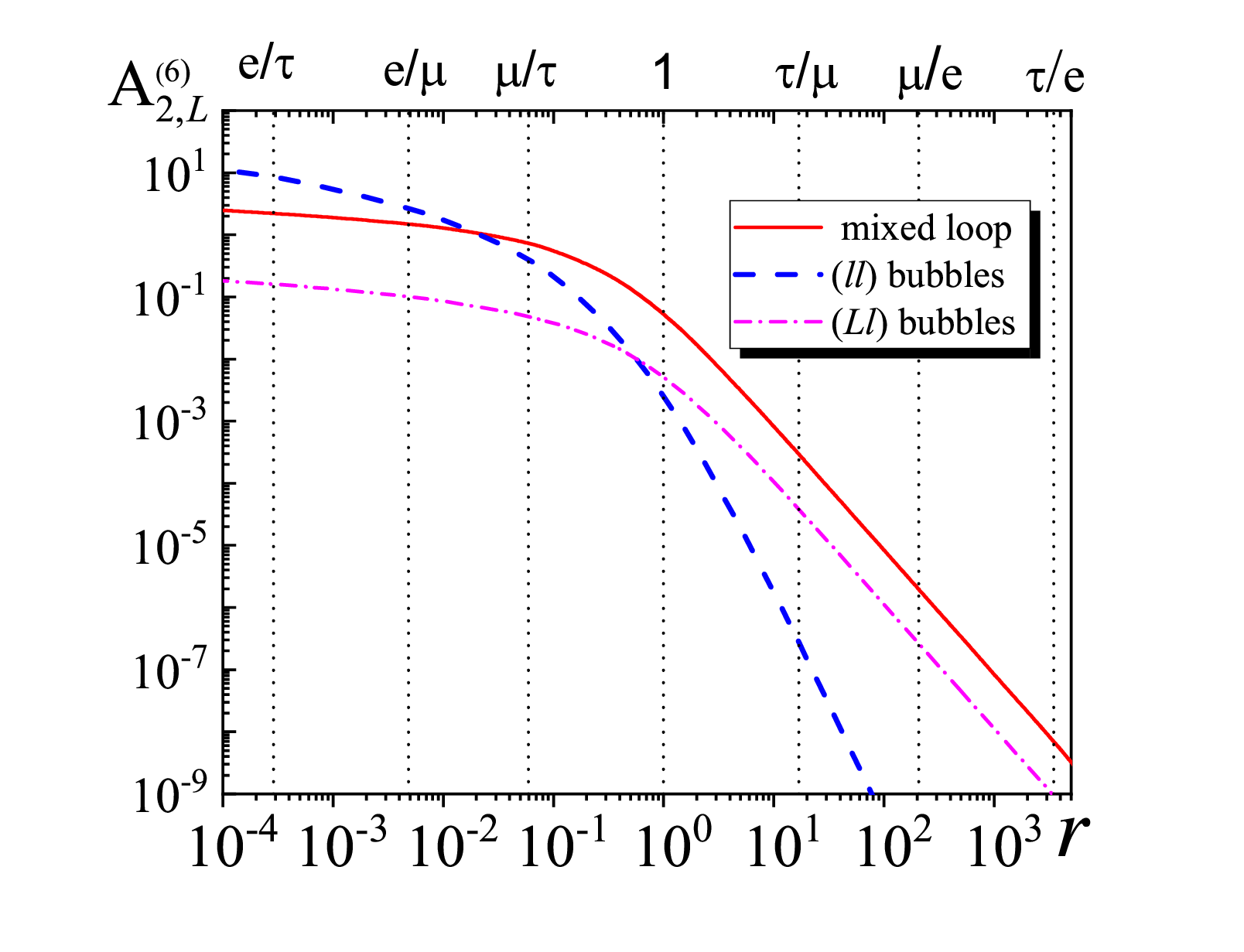}
\caption{ (color online) The sixth order corrections $A_{2,L}^{(6)}(r)$ to the lepton anomaly due to insertions of the fourth order polarization operator: the solid line  corresponds to  one loop   polarization operator with one internal photon line, the dashed and dot-dashed  lines  correspond  to insertions of the pure   bubble-like   operators with  two identical $(\ell\ell)$ and two different $(L\ell)$ closed lepton loops, respectively, cf. Ref.~\cite{Solovtsova23} and Appendix~\ref{appB}. The vertical lines, labeled according  to the ratios $ r= {m_\ell}/{m_L}$,  indicate the values $r$ of the physically existing leptons.
} \label{mixedAndbubbles}
 \end{figure}
This type of corrections was comprehensively  studied   in Ref.~\cite{Solovtsova23}, where
contributions of all possible combinations of internal and external types of leptons were investigated for polarization operators with two and three closed lepton loops
 within the Mellin-Barnes approach.
Explicitly, the sixth order coefficients $A_{2,L}^{(6)}$ due to the fourth-order bubble-like polarization operators are delegated to Appendix~\ref{appB}. In  Fig.~\ref{mixedAndbubbles}, the contribution of a different kind of  polarization operators is presented, where the solid curve is the correction $A_2^{(6)}(r)$ due to mixed diagrams, and dashed and dot-dashed lines are the corrections from the pure bubble-like polarization operators with insertions of two identical $(\ell\ell)$  loops and two loops with different  leptons  $(L\ell)$. In the region $r<1$ the main contribution to heavier  leptons, $\tau$ and $\mu$, comes from insertions of two identical electron loops. The contribution of  bubble-like diagrams with one heavy lepton  (dot-dashed line in Fig.~\ref{mixedAndbubbles}, $r<1$) is by far smaller than other contributions. The role of mixed loops increases with increase of $r$. A   different situation  is in the right semi-plane $r\ge 1$, where the mixed diagrams are predominant.

Finally, we briefly discuss the role of the remaining type of the sixth order bubble-like diagrams, namely, the ones with insertion of the polarization operators with all three  leptons  different from each other, $\ell_1\neq\ell_2\neq L$.  This corresponds to the coefficient  $A_{3,L}^{(6)}(r_1,r_2)$ in Eq.~(\ref{A4}). In Ref.~\cite{Friot:2005cu}, this  type of corrections was considered for  the muon anomaly with one  electron and one $\tau$ lepton loop  in the asymptotic limits $r_1=m_e/m_\mu \ll 1$ and $r_2=m_\tau/m_\mu \gg 1$. A more rigorous analysis of  $A_{3,L}^{(6)}(r_1,r_2)$ can be found in Ref.~\cite{Ananthanarayan:2020acj}, where  using the two-fold Mellin-Barnes representation, the earlier known asymptotic expansions for $A_{3,L}^{(6)}(r_1,r_2)$ were extended to their exact expressions  for all leptons~\footnote{It should be noted that in Ref.~\cite{Ananthanarayan:2020acj} the variables $r_1$ and $r_2$ are defined in an opposite way in comparison to ours, $r =m_L/m_\ell$.}, $L=e,\mu$~and~$\tau$.  It turns out that  the final analytical results are extremely  cumbersome    containing a number of  two-variable generalization of the  hypergeometric series, known as the Kamp\'{e} de F\'{e}riet functions. For this reason, the obtained analytical expressions were checked vs. the high precision calculations~\cite{LaportaPLB} to ensure that there is  agreement between integral and exact analytic results.
The reported numerical results are as follows:
\ba &&
A_{3,e}^{(6)}=1.90972\times 10^{-13}, \label{friot1} \\ &&
A_{3,\mu}^{(6)}=5.27737\times 10^{-4}, \label{friot2} \\ &&
A_{3,\tau}^{(6)}=3.34778. \label{friot3}
\ea
However, in practice one  can   avoid such sophisticated calculations by merely  performing direct
calculations of $A_{3,L}^{(6)}(r_1,r_2)$ from the general formula in Eq.~(\ref{double}).
In this case, the corresponding polarization operator can be written as
\ba
\widetilde\Pi\left( \frac{-x^2}{ 1-x }m_L^2\right)
 =&&2\left(\frac{\alpha}{\pi}\right)^2\left( \frac59 -\frac{4r_1^2}{3x^2} +\frac{4r_1^2}{3x}
 \sqrt{4r_1^2(1-x)+x^2} \ln{\frac{(x+\sqrt{4r_1^2(1-x)+x^2}^2}{4r_1^2(1-x)}} \right)\times\nonumber\\&&
 \left( \frac59 -\frac{4r_2^2}{3x^2} +\frac{4r_2^2}{3x}
 \sqrt{4r_2^2(1-x)+x^2} \ln{\frac{(x+\sqrt{4r_2^2(1-x)+x^2}^2}{4r_2^2(1-x)}} \right),
\label{pol12}
\ea
where the  explicit expression for the one-loop polarization operator $\Pi^{(\ell)}\left(\frac{-x^2}{1-x}m_L^2\right)$ was used,~Ref.~\cite{Aguilar:2008qj}. We employed Eq.~(\ref{pol12}) to calculate numerically the contribution to $A_{3,L}^{(6)}(r_1,r_2)$  from    Eq.~(\ref{double})~with insertion of the operator (\ref{pol12}). In our calculations   the variables $r_1$ and $r_2$ were used, as  already mentioned, from the CODATA2018~\cite{CODATA:2018}. We found that Eq.~(\ref{double}) exactly reproduces the  results presented in Ref.~\cite{Friot:2005cu}, cf. Eqs.~(\ref{friot1})-(\ref{friot3}).

A short glance at Fig.~\ref{mixedAndbubbles} and Eqs.~(\ref{friot1})-(\ref{friot3}) persuades us that there are regions where the corrections from one-loop  mixed diagrams $ A_{2,L}^{(6)}(r)$ and two-loop diagrams  $ A_{3,L}^{(6)}(r_1,r_2)$ are quite compatible with each other. For instance, the one-loop mixed coefficient $A_{2,\tau}^{(6) }(m_e/m_\tau)\approx 2.221259$    is of the same order of magnitude as the bubble-like coefficients  $A_{3,\tau}^{(6)}(m_e/m_\tau,m_\mu/m_\tau)$, Eq.~(\ref{friot3}), the coefficients $A_{2,\mu} (m_\tau/m_\mu)\approx 2.9474\cdot 10^{-4}$  are close to $A_{3,\mu}^{(6)}(m_e/m_\mu, m_\tau/m_\mu)$, Eq.~(\ref{friot2}). This is a clear evidence that in calculations of the corrections to $a_L$ from the vacuum polarization diagrams, all types of the fourth order polarization operators, the    bubble-like operators   ($\ell\ell$), ($L \ell $)  and   ($\ell_1\ell_2$), where  $\ell\neq L, \ell_1\neq L, \ell_2\neq L$, and the  mixed diagram (one lepton loop with an internal crossing photon line) equally contribute to on and must be considered alltogether on the same footing.

\section{Conclusions}\label{concl}
We have presented  for the first time  exact analytical expressions for the sixth order
radiative corrections to the anomalous magnetic moments of leptons $e$,~$\mu$~and~$\tau$
induced by  Feynman diagrams with insertions of the fourth order vacuum polarization operator with
one   lepton loop crossed by  one internal photon line. The approach essentially relies on
the dispersion relations and the Mellin-Barnes transform   for the propagators of massive photons
previously employed in calculations of the pure bubble-like diagrams. This method allows one  to
derive explicitly the corresponding sixth order corrections $a_L(r)$ as functions of the
 ratio $r=m_\ell/m_L$ of the mass of the internal $\ell$ to the mass of the external $L$ leptons
    in the whole interval ($0<r<\infty$).   It is argued that for each type of leptons the main
    contribution to $a_L(r)$ is provided by insertions of the polarization operator with leptons $\ell$
    in the loop lighter than the external lepton $L$, cf. labels  in Figs.~\ref{246012} and \ref{mixedAndbubbles}.
     Since for real  existing leptons one has either $r\ll 1$  $(r_{max} \lesssim 0.06$)
     or $r\gg 1\, (r_{min} \gtrsim 16)$, the exact expressions can be safely substituted
      by their asymptotic expansions, which are much simpler and more convenient for numerical calculations.
       We affirm  that these expansions work quite well in the intervals ($0\,< \,r\, <\, 0.1$) and  ($2\,< \,r\, <\,\infty$)
       within which the physical ratios $r=m_\ell/m_L$ are located. We investigated
       the limits of applicability of the asymptotic expansions and claim that with a
       reasonable number of terms these asymptotic are quite appropriate for practical numerical
       calculations. We estimated numerically the contribution to $a_L$ of diagrams with
       two lepton loops $\ell_1\neq L$ and $\ell_2\neq L$  by Eq.~(\ref{double}) and found perfect
       agreement with the results earlier reported in Ref.~\cite{Friot:2005cu}. We also compared
       the contribution to $a_L$ from all possible types of the fourth order polarization operator
       and argue that there are intervals where all operators equally contribute to $a_L$ and must
       be properly taken into account in calculations of the radiative corrections from   insertions
       of the vacuum polarization operators of the corresponding order.

       The Mellin-Barnes formalism is an extremely flexible technique that
        can be  straightforwardly generalized to analytical calculations of another class of diagrams   such as
        multi-loop  calculations  in relativistic quantum field theories~\cite{Kotikov:2018wxe}, evaluation
        of the contributions to $a_L$ of the hadronic vacuum polarization insertion~\cite{Charles:2017snx}, investigation of the physics
       Beyond the Standard Model, etc. Within the Mellin-Barnes approach,   it  is feasible to
        explicitly estimate   higher order  corrections from
         multi-mixed diagrams and
        combined types containing simultaneously  bubble-like and mixed loops  diagrams.
         We plan to discuss this in the near future.

\section{Acknowledgments}
 We highly appreciate valuable discussions with Prof. A.V.~Kotikov and V.V.~Bytev.
This work was supported in part by  a grant under the Belarus-JINR scientific collaboration. A bulk of numerical calculations was performed   on the basis of the HybriLIT heterogeneous computing platform~\cite{govorun} (supercomputer ``Govorun'', LIT, JINR).

\appendix
\section{} \label{appA}
Here we show that, as it should be,  the longitudinal terms $  \sim k_\mu k_\nu$ in the propagator (\ref{prop}) do not contribute to the lepton anomaly $a_L=(g_L-2)/2$. To this end, recall that the anomaly $a_L$ is just the normalization of the Pauli form factor $F_2(q^2)$ at $q^2=0$ in the general expression for the renormalized electro-magnetic lepton vertex $\Gamma_\mu(q)$, see e.g., Ref.~\cite{Solovtsova23}.
\begin{eqnarray} &&
 \bar u(p_2)   \Gamma_\mu(p_1,p_2)u(p_1) =\bar u(p_2) \left[ \gamma_\mu F_1(q^2) +
i \frac{\sigma_{\mu\nu} q^\nu}{2m_L} F_2(q^2)\right ] u(p_1)=\nonumber \\ &&
\bar u(p_2) \left[\frac{(p_1+p_2)^\mu}{2m_L}
F_1(q^2) + i \frac{\sigma_{\mu\nu} q^\nu}{2m_L} \left
(F_1(q^2)+F_2(q^2)\right)\right ]u(p_1),
\label{gamma1}
\end{eqnarray}
where   the second   raw in Eq.~(\ref{gamma1}) is a corollary of the Gordon identity for the on-shell fermions.
The form factor $F_2(0)$ in Eq.~(\ref{gamma1}) can be separated by applying on the vertex  $\Gamma_\mu(p_1,p_2)$
a properly defined projection operator, e.g.,  the projection operator  ${\cal P}_\mu$~\cite{Solovtsova23}
\ba
&&{\cal P}_{\mu} =  \frac{1}{Q^2}(\gamma_\mu \hat {q}  +p_\mu \hat{q}/m_L  - q_\mu )(\hat{p} + m_L)+
\frac13\gamma_\mu -\frac{p_\mu}{m_L}-\frac43 p_\mu \frac{\hat{p}}{m_L^2},\label{proj1}
\ea
where $p=(p_1+p_2)/2$ and $q=(p_2-p_1)$ with $(p\cdot q)=0$.
Then it is straightforward to show that
\ba
a_L=F_2(0)=\lim_{q\to 0}\left[ \frac14 Tr \left ({\cal P}^\mu \Gamma_\mu\right)\right ]. \label{trac}
\ea
\begin{figure}[h]
 \phantom{}\hspace*{-0.8cm}%
\includegraphics[width=0.35\textwidth,clip=true]{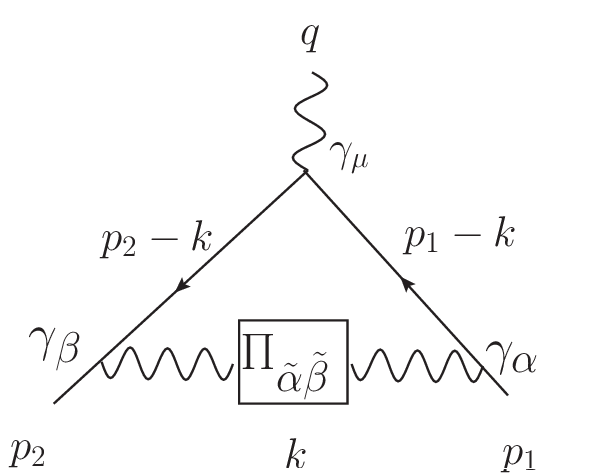}
\caption{A schematic illustration of the Feynman diagrams with vacuum
polarization insertions.  The operator
  $\Pi_{\tilde\alpha\tilde\beta}(k^2)$
   is transverse,
   $\Pi_{\tilde \alpha\tilde\beta}(k^2)=\left[ g_{\tilde\alpha\tilde\beta} - k_{\tilde\alpha} k_{\tilde\beta}/k^2 \right]\Pi(k^2)$
    so that the gauge term $(1-\xi) \sim k_\mu k_\nu$ in the dressed photon propagator (\ref{prop}) keeps the product
   $D_{\alpha\tilde\alpha}(k^2) \Pi_{\tilde\alpha\tilde\beta}(k^2) D_{ \tilde\beta\beta}(k^2)$ transverse
  $D_{\alpha\beta}(k^2) =\left( g_{\alpha\beta}-k_{\alpha} k_{\beta}/k^2 \right )
   \frac{1}{k^2} \dfrac{1}{\big(1+      \Pi (k^2) \big )}
$. }
\label{gauge}
 \end{figure}

In our case, the corresponding vertex   $\Gamma_\mu(q)$ is defined by the  Feynman diagram contributing to the lepton anomaly as depicted in Fig.~\ref{gauge}, where
$\Pi_{\tilde\alpha\tilde\beta}(k^2)$ is the vacuum polarization operator, which is transverse with respect to $k$, i.e.
$k_{\tilde\alpha} \Pi_{\tilde\alpha\tilde\beta}(k^2)= k_{\tilde\beta} \Pi_{\tilde\alpha\tilde\beta}(k^2)=0$.

 Then the contribution of the longitudinal part $\sim k_\alpha k_\beta$ of the propagator (\ref{prop}) reads as
 \begin{eqnarray} &&
 \bar u(p_2) \Gamma_\mu^{long.} (p_1,p_2)u(p_1) \simeq
 \bar u(p_2) \hat k (\hat p_1 -\hat k +m_L) \gamma_\mu ( \hat p_2 -\hat k +m_L) \hat k \bar u(p_1) = \nonumber\\ && =
 4\big( (p_2\cdot k) -k^2\big) \big((p_1\cdot k) -k^2\big) \bar u(p_2) \gamma_\mu u(p_1) =\nonumber\\ && =
 4\big( (p_2\cdot k) -k^2\big) \big((p_1\cdot k) -k^2\big)  \bar u(p_2)  \gamma^\mu  u(p_1)
\frac{1}{2m_L} \bar u(p_2)\left
[p^\mu+i\sigma^{\mu\nu}q_\nu\right] u(p_1),
  \label{d1}
\end{eqnarray}
where the on-shellness   conditions $(\hat p_1-m_L)u(p_1)=\bar u(p_2)(p_2-m_L)=0$ and the Gordon identity have been
used.
Direct calculations of  the traces  in (\ref{trac}) with (\ref{proj1}) and (\ref{d1})   show that the contribution of $\Gamma_\mu^{long.}$
to $ F_2(0)$, Eq.~(\ref{trac}), is exactly zero.
It implies that in concrete calculations one can safely use the ``simplified'' photon propagator~\cite{Solovtsova23,lautrupNuclPhys,Aguilar:2008qj} neglecting all terms~$\sim k_{\alpha}k_{\beta}$ in Eq.~(\ref{prop}),
\ba && \
D_{\alpha\beta}(k^2) =  -i\left\{
\frac{ g_{\alpha\beta}}{k^2} \dfrac{1}{1+     \mathlarger{ \mathlarger{ \Pi (k^2)}}} \right\}.
\ea

\section{Contributios of the bubble-like two-loops diagrams}  \label{appB}

For the sake of completeness,  here below we briefly recall the results of calculations of the sixth order corrections to the anomaly $a_L$ due to the insertion of the fourth order bubble-like vacuum polarization operators. In Ref.~\cite{Solovtsova23}, it was shown that each of  the corresponding coefficients $A_{2,L}^{(6),(\ell_1 \ell_2) }(r<1)$  and $A_{2,L}^{(6),(\ell_1 \ell_2) }(r>1)$, where the superscripts  denote the loops formed by the leptons $(\ell_1 \ell_2)$, can be analytically continued  by the corresponding  single  analytical function of $r$, valid in the whole interval $r \in (0,\infty)$.

\begin{figure}[!hbt]
\vspace*{4mm}
\includegraphics[width=0.3\textwidth]{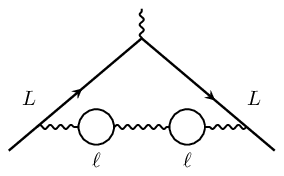}
\hspace*{16mm}
\includegraphics[width=0.3\textwidth]{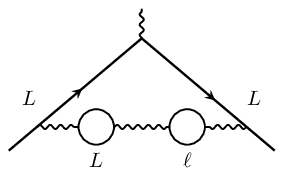}
\caption{  Feynman diagrams of the bubble-like type with two identical (left panel)
  and   two different (right panel)  lepton loops contributing
to the sixth order radiative corrections to the anomaly $a_L$.} \label{Fig-2loop}
\end{figure}

For the $(\ell \ell)$ diagrams, corresponding to the left panel in Fig.~\ref{Fig-2loop},
the sought function is
\ba &&
 A_{2,L}^{(6),(l l) }(r)=\frac{2}{3} \left({\frac{1}{3}}-4\,
r^2 + 5\, r^4 - {\frac{16}{15}}\, r^6 \right)
\left[-{\rm
Li_2} \left( \frac{1-r}{1+r} \right)+{\rm Li_2} \left( -
\frac{1-r}{1+r} \right) -2{\rm Li_2}
\left(1-\frac{1}{r} \right)
  \right.\nonumber \\ &&
 \left.
 + \frac{1}{12}\;\pi^2 \right] \,
- \frac{8}{45}\left[{\rm Li_2}\left(\frac{1-r}{1+r} \right)-{\rm Li_2}\left(- \frac{1-r}{1+r} \right)\right.
 \left.
+\frac{1}{4}\;\pi^2\right] r 
+\frac{8}{3}\,\left[{\rm
Li_3}\,({r}^{2})- \bigg( {\rm Li_2}\,
({r}^{2}) +\frac{1}{3}\;\pi^2 \bigg)\ln (r)
\right] r^4
   \nonumber \\ &&
 + \frac{317}{324}  -\frac{191}{45}\;r^2
+\left(\frac{25}{27} -\frac{254}{45}\,r^2+ \frac{32}{45} \, r^4 \right)\ln(r)
%
+
\frac{16}{45}\, r^4
-\frac{16}{9} \, r^4 \ln^3 (r).
 \label{A2eq}
 \ea

For the diagrams with two different lepton loops $(L\neq \ell)$, right panel in Fig.~\ref{Fig-2loop},
the corresponding  function  $A_{2,L}^{(6),( L\ell) }(r)$ is
\begin{eqnarray}
  && A_{2,L}^{(6),( L\ell) }(r)=
 \left[ \frac {4}{45\,{r}^{2}}  +\frac19+\frac43\,{r}^{2} - \frac {13}{9}\,r^4 \right]
    \bigg[ -{\rm Li_2} \left( 1-\frac{1}{{r}^{2}} \right) +\frac{\pi^2}{6}\bigg]- 2\left( 1+{r}^{4}\right) \times
   \nonumber
\\ &&
     {\rm Li_3} \left( \frac{1}{{r}^{2}} \right) +
\bigg( \frac1r+\frac23\,r+\frac{11}{3}\,r^3  \bigg)
\bigg\{
 \bigg[{\rm Li}_2
\left(\frac{1}{r}\right)-{\rm Li}_2
\left(-\frac{1}{r} \right)
 \bigg] \ln(r)
 +\frac12\big[ \ln(1+r)  - \ln(r-1) \big]
  \nonumber
\\ &&
\times
 \ln^2(r)+{\rm Li}_3
\left(\frac{1}{r}\right)-{\rm Li}_3
\left(-\frac{1}{r} \right) \bigg\}
    - \frac {8}{3}\;(1+r^4)
  \bigg[ {\rm Li_2} \left( \frac{1}{{r^2}} \right) - \frac12 \ln\left(1-\frac{1}{r^2} \right)\ln(r)\bigg]\ln(r)
   \nonumber \\ &&
 +\frac {8}{45}\;r^3
 \left[{\rm Li_2}\left(\frac{r-1}{r+1} \right)-{\rm
Li_2}\left(- \frac{r-1}{r+1} \right)
+\frac{1}{4}\;\pi^2\right]
-  \bigg(\frac {8}{45 r^2}+\frac{11}{9}+\frac{7}{3}r^2 \bigg)\ln^2(r)
 \nonumber
\\ &&
-\frac{
1853}{810} -\frac{349}{135} \ln(r) -\frac {53}{15} \, r^2 - \frac{64}{45}\,r^2 \ln(r) -{\frac {4\,{\pi}^{2}}{135\,{r}^{2}}} \, .
  \label{lLL}
\end{eqnarray}

 The dependence of  the coefficients $A_{2,L}^{(6),(\ell\ell)}(r)$, Eq.~(\ref{A2eq}), and $A_{2,L}^{(6),(L\ell)}(r)$,   Eq.~(\ref{lLL}),
   vs. $r$ is presented  in Fig.~\ref{mixedAndbubbles}, the dashed and dot-dashed lines, respectively.

\end{document}